\documentclass[aps,preprint,showpacs,preprintnumbers,amsmath,amssymb]{revtex4}

\usepackage{graphicx}
\usepackage[english]{babel}

\begin{document}

\def \B{\mathcal{B}}
\def \A{\mathcal{A}}
\def \Z{\mathcal{Z}}
\def \Tr{\mathrm{Tr}\,}
\def \Re{\mathrm{Re}\,}
\def \Im{\mathrm{Im}\,}
\def \intk{\nu \int \frac{\mathrm{d}^3k}{(2\pi)^3} \,}
\def \intw{\int_{-\infty}^{\infty} \frac{\mathrm{d} \omega}{2\pi} \,}
\def \eqp{\varepsilon_{qp}(k)}
\def \ebhf{\varepsilon_{BHF}(k)}
\def \ek{\frac{k^2}{2m}}

\title{The entropy of a correlated system of nucleons}

\author{A. Rios, A. Polls and A. Ramos}

\affiliation{Departament d'Estructura i Constituents de la Mat\`eria,
         Universitat de Barcelona, Avda. Diagonal 647, E-08028 Barcelona, Spain}

\author{H. M\"uther}

\affiliation{Institut f\"ur Theoretische Physik, \\
		 Universit\"at T\"ubingen,\\
		 D-72076 T\"ubingen, Germany}

\begin{abstract}
Realistic nucleon-nucleon interaction induce correlations to the nuclear
many-body system, which lead to a fragmentation of the single-particle strength
over a wide range of energies and momenta. We address the question of how this
fragmentation affects the thermodynamical properties of nuclear matter. In
particular, we show that the entropy can be computed with the help of a spectral
function, which can be evaluated in terms of the self-energy obtained in the
Self-Consistent Green's Function  approach. Results for the density and
temperature dependences of the entropy per particle for symmetric nuclear matter
are presented and compared to the results of lowest
order finite temperature Brueckner--Hartree--Fock calculations. The effects of
correlations on the calculated entropy are small, if the appropriate
quasi-particle approximation is used. The results demonstrate the thermodynamical
consistency of the self-consistent $T$-matrix approximation for the evaluation of
the Green's functions.
\end{abstract}


\maketitle

\section{Introduction}

The entropy is a key quantity in the study of the thermodynamical (TD) properties of
fermionic systems. In the context of correlated Fermi liquids, the study of the
entropy has been triggered by the experimental and theoretical studies of $^3$He 
\cite{berk66,doniach66,amit68,brenig67,carneiro73,carneiro75}. The specific heat of this system
is experimentally known to have a non-trivial temperature dependence \cite{greywall83}
of the type $T^3 \ln T$. Such a non-analytical behaviour can be seen to arise
within Fermi liquid theory from the coupling between quasi-particles and incoherent
spin fluctuations (quasi-holes in the triplet state), which give rise to non-analytical
energy dependences in the self-energy \cite{chubukov06}. These non-analyticities are in
fact a very general feature of all normal Fermi liquids and their existence is not
related to the particular details of $^3$He systems .

Hot symmetric nuclear matter (SNM), on the other hand, is an infinite fermionic system
composed of nucleons at very high densities. In this ideal system, only the strong
interaction among nucleons is considered and any other interaction, such
as the electromagnetic one between protons, is neglected.
The temperatures usually considered for this system are of the order of tenths
of MeV and are somewhat small if compared with the typical energy scales of nuclear matter
(the free Fermi energy at saturation density is about $\epsilon_F \sim 40$ MeV, thus
$T/\epsilon_F \sim 0.5$ for $T=20$ MeV, the highest temperatures considered here).
We expect that a finite temperature approach to SNM can fairly describe the properties
of the hot environments that could
exist either inside the cores of supernovae at the latest stage of their evolution
\cite{bethe90} or in the collisions of heavy nuclei at intermediate energies
\cite{chomaz04}. The TD properties of these systems and, in particular,
the entropy are important quantities for the understanding of astrophysical and heavy
ion physics phenomena. 
In core-collapse supernovae, for instance, the evolution and dynamics occur at a
fixed entropy per baryon \cite{prakash97}. Moreover, the entropy production in
multi-fragmentation events in heavy ion collisions is considered to be a crucial
quantity to determine the mass fragment distribution \cite{csernai86}.

Hot SNM has been usually studied in a mean-field approximation with
effective phenomenological nucleon-nucleon (NN) forces, such as the Skyrme or
Gogny interactions \cite{sil04,heyer88}, or within a Relativistic Mean Field
approximation
\cite{muller95}. Other many-body approaches that have been used in the study
of SNM include lattice models \cite{muller00} or three-loop calculations
within chiral perturbation theory at finite temperature \cite{fritsch02}.
However, when dealing with realistic NN potentials, more sophisticated
many-body techniques are needed. The strong short range repulsion
and the tensor components of these potentials modify substantially the many-body
wave function, which is not anymore well described in terms of a free Fermi gas
Slater determinant.
Particle-particle correlations, for instance, are crucial for a correct
description of SNM properties from realistic NN potentials. The 
Brueckner--Hartree--Fock (BHF) approach accounts for such correlations by means
of a summation of an infinite series of suitable ladder diagrams \cite{brueckner54}.
In fact, the BHF approximation arises from a well-defined expansion for the
energy of a fermionic system at zero temperature, the so-called
Brueckner--Bethe--Goldstone expansion \cite{day67}. At finite temperature,
a similar summation can be achieved in the so-called
Bloch--de Dominicis (BdD) approach \cite{bloch58}. However, this approach is
no devised to reproduce the energy of the system. Instead, it aims to compute
the TD grand-potential and, from it, all the TD properties
of the system. In particular, the entropy includes the correlations embedded in
the approach \cite{baldo99}. 

Traditionally, however, the BHF approach has
been extended to finite temperatures in a more naive way: the energy of the system
is computed from a simple generalization of the $T=0$ formalism to finite
temperature and the entropy of the system is computed from a mean-field expression
\cite{lejeune86,bombaci94}.
Finally, let us also note that relativistic BHF-type calculation at finite
temperature have also been performed in the literature \cite{terhaar86,huber98}.

A consistent treatment of correlations in quantum many-body systems
requires the inclusion of particle-particle and hole-hole scattering terms. 
The usual approach is the
so-called ladder approximation or $T$-matrix approximation, in which an 
expansion of the single-particle
Green's function in terms of diagrams is achieved. Such an expansion goes beyond the
BHF approach in the sense that both the propagation of particles and
holes are treated in the same footing. Within this approach, the
propagator of a correlated fermion is not simply described in terms of one
quasi-particle pole.
Instead, the strength of a momentum state {\it k} is fragmented over a wide
range of energies. The process of finding the Green's function for such a many-body
system is unavoidably self-consistent, since the propagation of a nucleon is affected
by the interactions with the surrounding nucleons, which in turn are also described
in terms of Green's function. Although one can easily write down the diagrammatic expansion
that gives rise to this Self-Consistent Green's Function (SCGF), it has taken
a lot of time to have a numerical solution of the full problem. A major well known issue
is related to the so-called pairing instability that appears in the zero temperature 
formalism when the propagation of holes is considered \cite{vonderfecht93,alm96,bozek99}.
Nevertheless, calculations have been performed at the SCGF level at zero
temperature within different approximations: with a quasi-particle self-consistent approach
\cite{ramos89}, with a discretized parameterization of the spectral function \cite{dewulf02}
or with a simplified separable NN interaction \cite{bozek02}. In all these approaches the final
output is the single-particle spectral function of a nucleon inside matter, which can also be
obtained from other many-body approaches, such as the Correlated Basis Function
theory \cite{benhar89}.

In the case of SCGF, a finite temperature treatment has been the keystone for
obtaining a complete numerical solution from a realistic NN potential
\cite{frick03,frick04,frick05}. Once the self-consistent propagator is obtained in this
method, one can easily obtain information on the microscopic properties (such as
the momentum distributions, self-energies or spectral functions of the nucleon)
as well as on the bulk properties of the system (the energy per
particle via the Galitski-Migdal-Koltun (GMK) sum rule, for instance
\cite{migdal58,koltun74}). For a complete TD
description of the system, however, one should compute the relevant TD potential
of a statistical quantum mechanical system, \emph{i.e.} the free energy. A
suitable calculation of the entropy is thus required if this formalism is to be
used in any practical description of hot SNM. Here, we will follow the Luttinger-Ward
(LW) approach \cite{carneiro75,luttinger60}, in which the grand-potential is computed
from the full single-particle propagator.
An analysis of the properties of SNM within this formalism has been recently published
by Soma \emph{et al.} \cite{soma06}. In the following we shall show
that one does not need to compute TD quantities within a full LW approach as done in
\cite{soma06}, provided that some approximations for the entropy are valid.
Finally, let us notice that the LW formalism has been widely used in other
many-body physics problems, ranging from relativistic plasmas \cite{vanderheyden98} to
resonances in heavy ion collisions \cite{weinhold98}.

In the following section, we will describe in detail the SCGF approach at finite
temperature and the LW formalism that we will use in our calculations
of the entropy. The numerical results derived from this formalism
will be divided in two different parts. Section III will be devoted to the
microscopic results, while Section IV will describe the bulk TD properties of a
correlated system of nucleons. Finally, a brief summary will be given in Section V.

\section{Formalism}

\subsection{The ladder approximation}

With the help of the single-particle propagator, we can obtain all the one-body
(and even some two-body) properties of a many-body system \cite{fetter71}. In the Green's function
approach, one aims to compute the single-particle propagator which, in the
grand-canonical ensemble is defined according to:
\begin{eqnarray}
i G(\mathbf{k}t,\mathbf{k}'t') &=& \Tr \Big\{ \hat{\rho} \, \mathcal{T} \big[ a_{\mathbf{k}}^{}(t) a^{\dagger}_{\mathbf{k}'}(t') \big] \Big\}	\, ,	
\label{eq:spprop}
\end{eqnarray}
where we have introduced the density matrix operator:
\begin{eqnarray}
\hat{\rho}=\frac{1}{Z} e^{-\beta(\hat{H}-\mu \hat{N})} \, ,
\label{eq:denmat}
\end{eqnarray}
and the partition function:
\begin{eqnarray}
Z=\textrm{Tr} \Big\{ e^{-\beta(\hat{H}-\mu \hat{N})} \Big\} \, .
\label{eq:partfunct}
\end{eqnarray}
In Eq.~(\ref{eq:spprop}), $\mathcal{T}$ is the time-ordering operator in such a 
way that the Heisenberg creation (annihilation) operator $a_{\mathbf{k}}^{\dagger}(t)$ 
($a_{\mathbf{k}}^{}(t)$) with the largest time argument $t$ (or $it$ in the
case that $t$ is imaginary) is put to the left, with a minus sign included for each
commutation. In these equations $\beta$ denotes the inverse temperature and 
$\mu$, the chemical potential of the system. For simplicity, we will neglect the
spin-isospin structure of this propagator in the following. Finally, the traces
$\Tr$ are to be taken over all the energy and particle number eigenstates of the
system. The cyclic invariance of these traces imply the following quasiperiodicity
condition for the Green's function:
\begin{eqnarray}
G(\mathbf{k}t=0,\mathbf{k}'t') &=& - e^{\beta \mu} G(\mathbf{k}t=-i \beta,\mathbf{k}'t')	\, .
\label{eq:qpcond}
\end{eqnarray}
One can thus Fourier transform the time dependence of the propagator in terms
of some Fourier coefficients $G(k,z_{\nu})$, depending on the discrete Matsubara frequencies
$z_{\nu} = \frac{(2 \nu + 1) \pi}{- i \beta} + \mu$. A Lehmann decomposition of these coefficients
is achieved by means of the spectral function $\A(k,\omega)$:
\begin{eqnarray}
G(k,z_{\nu}) &=& \intw \frac{\A(k,\omega)}{z_{\nu}-\omega}	\, .
\label{eq:lehmann}
\end{eqnarray}
The function $G(k,z_{\nu})$ evaluated at the Matsubara frequencies can be analytically
continued for all non-real $z$. Using the Plemelj formula, the spectral function can be
related with the values of $G$ close to the real axis (we use the notation $\omega_+=\omega+i \eta$):
\begin{eqnarray}
\A(k,\omega) &=& -2 \, \Im G(k,\omega_+)	\, .
\label{eq:asf}
\end{eqnarray}

The single-particle Green's function can be obtained from Dyson's equation, which for
a translationally invariant system reduces to the algebraic equation:
\begin{eqnarray}
\bigg[ \omega - \ek - \Sigma(k,\omega) \bigg] G(k,\omega) &=& 1	\, ,
\label{eq:dyson}
\end{eqnarray}
where $\Sigma(k,\omega)$ denotes a complex self energy. 
By expanding the self-energy in terms of one-particle Green's function, it can be
shown that $\Sigma$ and $G$ share the same analytic properties. Thus, one writes the
following spectral decomposition for $\Sigma$:
\begin{eqnarray}
\Sigma(k,z) &=& \Sigma^{HF}(k) + \intw \frac{\Gamma(k,\omega)}{z-\omega}	\, ,
\label{eq:sigma_dec}
\end{eqnarray}
where the width $\Gamma(k,\omega)$ is related to the imaginary part of $\Sigma$ by:
\begin{eqnarray}
\Gamma(k,\omega) = -2 \, \Im \Sigma(k,\omega_+) \, ,
\label{eq:width}
\end{eqnarray}
and it is real and positive for all energy and momenta \cite{luttinger61}. The
term $\Sigma^{HF}(k)$ in Eq.~(\ref{eq:sigma_dec}) is a real energy-independent
generalized Hartree-Fock contribution to the self-energy.

The self-energy can be derived from an in-medium two-body interaction (the so-called 
scattering $T$-matrix) which includes the correlations induced by the strong
short-range and tensor components of a realistic two-body NN force. Within the
ladder approximation, one can indeed relate the self-energy and the retarded
$T$-matrix by \cite{kadanoff62,kraeft86}:
\begin{eqnarray}
\textrm{Im }\Sigma(k,\omega_+) &=&
							\int \frac{\textrm{d}^3 k'}{(2\pi)^3}
							\int_{-\infty}^{\infty} \frac{\textrm{d} \omega'}{2\pi}
							\langle \mathbf{kk}' | \textrm{Im }T(\omega+\omega'_+) | \mathbf{kk}' \rangle_A
							 \A(k',\omega') \nonumber \\
							&\times& \big[ f(\omega')+b(\omega+\omega')\big] \, ,
\label{eq:sigma_ladder}
\end{eqnarray}
where we have introduced the Fermi-Dirac:
\begin{eqnarray}
f(\omega) = \frac{1}{e^{\beta [\omega - \mu]} + 1} \, ,
\label{eq:FD}
\end{eqnarray}
and the Bose-Einstein:
\begin{eqnarray}
b(\Omega) = \frac{1}{e^{\beta [\Omega - 2 \mu]} - 1}
\label{eq:BE}
\end{eqnarray}
distributions. The pole of the Bose function $b(\Omega)$ at $\Omega = 2 \mu$ is 
exactly canceled by a zero in the $T$-matrix \cite{alm96,alm93} and thus the integrand
remains finite as long as the $T$-matrix does not acquire a pole at this energy.
This pole appears only for temperatures below a certain critical temperature
and it is closely related to the onset of pairing among nucleons
\cite{alm96,sedrakian95}.

\begin{figure}[thb]
   \includegraphics[height=3cm]{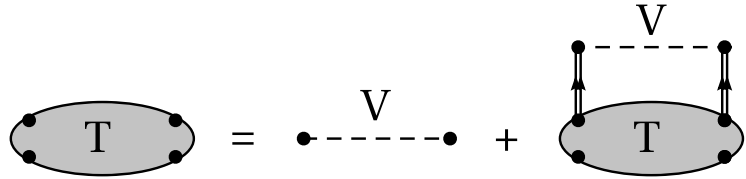}
   \vspace{0.75cm}
   \caption{Diagrammatical representation of the $T$-matrix within the ladder approximation.}
   \label{diag:tmat}
\end{figure}

Within the ladder approximation, the in-medium $T$-matrix is determined as a
solution of the integral equation:
\begin{eqnarray} 
\langle \mathbf{k k'} | T (\Omega_+) | \mathbf{p p'} \rangle_A &=&
				\langle \mathbf{k k'} | V | \mathbf{p p'} \rangle_A \nonumber \\
				&+& \int \frac{\textrm{d}^3 q}{(2\pi)^3} \, \int \frac{\textrm{d}^3 q'}{(2\pi)^3} \, 
				\langle \mathbf{k k'} | V | \mathbf{q q'} \rangle_A \,
				G^0_{II}(\mathbf{q q'},\Omega_+) \nonumber \\
				&\times& \langle \mathbf{q q'} | T (\Omega_+) | \mathbf{p p'} \rangle_A \, ,
\label{eq:tmat_sc}
\end{eqnarray}
where we have introduced the two-particle Green's function of two non-interacting but 
dressed nucleons:
\begin{eqnarray} 
G_{II}^0(k_1,k_2,\Omega_+)&=& 
				\int_{-\infty}^{\infty} \frac{\textrm{d} \omega}{2\pi} \int_{-\infty}^{\infty} \frac{\textrm{d} \omega'}{2\pi}
				\, \A(k_1,\omega) \A(k_2,\omega') \frac{1 - f(\omega) - f(\omega')}{\Omega_+-\omega-\omega'} \, .
\label{eq:gII0}
\end{eqnarray}
Diagrammatically, the $T$-matrix approximation is depicted in Fig.~\ref{diag:tmat}, where the
dressed one-particle propagators are given by two double lines. 
In general, the equations of motion arising within the Green's function method
couple the $N$-particle propagator to the $N+1$-particle propagator whenever we have
a two-body interaction. The Dyson equation for the single-particle Green's function 
involves, for instance, the full two-particle Green's function $G_{II}$. In the 
Self-Consistent Green's Function approach within the ladder approximation, however,
one ignores the effects of
three- and more particle propagators by using a function, $G^0_{II}$,
which is a product of two dressed one-body propagators. In that way, one solves
self-consistently the equations for the one-body $G$ and the approximated two-body
$G^0_{II}$ propagators.

If one wants to solve Eq.~(\ref{eq:tmat_sc}) in an efficient way, a partial wave
decomposition is needed. After expressing $G_{II}^0$ as a function of the total
$\mathbf{P}=\mathbf{k_1+k_2}$ and the relative momenta
$\mathbf{k_r}=\mathbf{(k_1-k_2)}/2$, one can perform the usual angle-average
approximation on the angle between $\mathbf{P}$ and $\mathbf{k_r}$, so that $G_{II}^0$
is expressed only in terms of the modulus $P$ and $k_r$:
\begin{eqnarray}
\bar G^0_{II}(P,k_r,\Omega_+) = \frac{1}{2} \int_{-1}^{1} \textrm{d} (\cos \theta) \,
G^0_{II}\left( \left| \mathbf{P + k_r} \right|, \left| \mathbf{P - k_r} \right|, \Omega_+ \right) \, .
\label{eq:gII0_av}
\end{eqnarray}
This approximation leads to a decoupling of the partial waves with different
total angular momentum $J$, which in its turn implies that Eq.~(\ref{eq:sigma_ladder})
becomes a one-dimensional integral equation:
\begin{eqnarray} 
\langle k_r | T^{JST}_{LL'} (P,\Omega_+) | k_r' \rangle_A &=&
                        \langle k_r | V^{JST}_{LL'} | k_r' \rangle_A \nonumber \\
                        &+& \sum_{L''} \int_0^{\infty} \frac{\textrm{d} k_r''}{(2 \pi)^3}
                         \, k_r''^2 \langle k_r | V^{JST}_{LL''} | k_r'' \rangle_A \,
                        \bar{G}^0_{II}(P,k_r'',\Omega_+) \nonumber \\
                        &\times& \langle k_r'' | T^{JST}_{LL''}(P,\Omega_+) | k_r' \rangle_A \, ,
\label{eq:tmat_pw}
\end{eqnarray}
By summing over all partial waves, we get the $T$-matrix as needed in the solution
of Eq.~(\ref{eq:tmat_sc}):
\begin{eqnarray} 
\langle \mathbf{k k'} | \Im T(\Omega_+) | \mathbf{k k'} \rangle_A &=&
                        \frac{1}{4 \pi} \sum_{JSTL} (2J+1) (2T+1) \nonumber \\
                        &\times&
                        \langle q(\mathbf{k,k'}) | \Im T^{JST}_{LL}(P(\mathbf{k,k'}),\Omega_+) | q(\mathbf{k,k'}) \rangle_A \, .
\label{eq:tmat_sum_pw}
\end{eqnarray}
The only remaining piece is now the generalized Hartree-Fock contribution to
the self-energy:
\begin{eqnarray} 
\Sigma^{HF}(k) = \frac{1}{8 \pi} \sum_{JSTL} (2J+1) (2T+1)
\int \frac{\textrm{d}^3 k'}{(2\pi)^3}
\langle q(\mathbf{k,k'}) | V^{JST}_{LL} | q(\mathbf{k,k'}) \rangle_A \, n(k') \, ,
\label{eq:HF_self}
\end{eqnarray}
where we have introduced the momentum distribution:
\begin{eqnarray} 
n(k) = \intw \A(k,\omega) f(\omega) \, .
\label{eq:nk}
\end{eqnarray}

In a self-consistent procedure, Eqs.~(\ref{eq:sigma_dec}), (\ref{eq:sigma_ladder}),
(\ref{eq:tmat_sc})-(\ref{eq:tmat_pw}) are solved at a given temperature and
density. The chemical potential $\mu$ is determined at each iteration by inverting:
\begin{eqnarray} 
\rho = \intk n(k,\mu) \, ,
\label{eq:den}
\end{eqnarray}
where $\nu$ denotes the spin-isospin degeneracy of the system ($\nu=4$ in the case
symmetric nuclear matter).
As a final output, once convergence is reached, one obtains the single-particle
propagator from Dyson's equation and, from its imaginary part, the spectral
function $\A(k,\omega)$. From this function, several micro- and
macroscopic properties of the system can be computed. For instance, the momentum
distribution given by Eq.~(\ref{eq:nk}) or the total energy per particle of the system,
accessible from the GMK sum rule:
\begin{eqnarray} 
\frac{E^{GMK}}{A} = \frac{\nu}{\rho} \int \frac{\mathrm{d}^3k}{(2\pi)^3} \intw
\frac{1}{2} \left( \ek + \omega \right) \A(k,\omega) f(\omega) \, ,
\label{eq:koltun}
\end{eqnarray}
where $A$ is the total number of particles of the system. 
Further on, we will assess the question of how to compute the partition
function and the entropy of a system of nucleons from the spectral function
$\A$.

Before doing so, however, let us consider some interesting approximations to the
SCGF method. Using Dyson's equation, it is straightforward to show that the
spectral function can be written as:
\begin{eqnarray}
\A(k,\omega) &=&	\frac{\Gamma(k,\omega)}{\left[ \omega - \ek - \Re \Sigma(k,\omega) \right]^2 +
                \left[ \frac{\Gamma(k,\omega)}{2} \right]^2 } \, .
\label{eq:asf_lor}
\end{eqnarray}
It is interesting to note that this is a positive defined function for all energy
and momenta \cite{luttinger61}, fulfilling the following sum rule
\cite{polls94,frick04b,rios06}:
\begin{eqnarray}
\intw \A(k,\omega) &=& 1 .
\label{eq:a_sumrule}
\end{eqnarray}
One can obtain a simplified set of equations by taking the zero-width limit, $\Gamma \to 0$ in
the previous expression, thus obtaining the $\delta$-function:
\begin{eqnarray}
\A(k,\omega) = 2 \pi \, \delta [\omega - \eqp] \, ,
\label{eq:asf_qp}
\end{eqnarray}
where the quasi-particle spectrum is derived by means of the real part of the
self-energy:
\begin{eqnarray}
\eqp = \ek + \Re \Sigma(k,\eqp) \, .
\label{eq:qp}
\end{eqnarray}
This is the so-called Quasi-Particle Self-Consistent Green's Function method.
In the following, we will not use this approximation. However, we shall extensively
use the $\delta$-peak approximation to the spectral function, Eq.~(\ref{eq:asf_qp}),
with the single-particle energies $\eqp$ defined by Eq.~(\ref{eq:qp}) with the
self-energy of our SCGF calculation.

In addition, we will also consider the BHF approach to nuclear
matter. This can be derived from the SCGF method if one takes the quasi-particle
approximation, Eq.~(\ref{eq:asf_qp}), and, in addition, one makes the following
substitution:
\begin{eqnarray}
(1 - f(\omega) - f(\omega') ) \to [1 - f(\omega)][1- f(\omega')]
\label{eq:BHF_GII}
\end{eqnarray}
in the two particle propagator, Eq.~(\ref{eq:gII0}). In the $T \to 0$ limit,
this expression becomes the Pauli operator and $G_{II}^0$ reduces to the
particle-particle propagator. From this approximation, one can see that the $T$-matrix
reduces to the well-known $G$-matrix in-medium interaction:
\begin{eqnarray} 
\langle \mathbf{k k'} | G (\Omega_+) | \mathbf{p p'} \rangle_A &=&
                        \langle \mathbf{k k'} | V | \mathbf{p p'} \rangle_A \nonumber \\
                        &+& \int \frac{\textrm{d}^3 q}{(2\pi)^3} \, \int \frac{\textrm{d}^3 q'}{(2\pi)^3} \, 
                        \langle \mathbf{k k'} | V | \mathbf{q q'} \rangle_A \,
                        \frac{[ 1 - f[\varepsilon_{BHF}(q)]][1- f[\varepsilon_{BHF}(q')]]}
                        {\Omega_+ - \varepsilon_{BHF}(q) - \varepsilon_{BHF}(q')} \nonumber \\
                        &\times& \langle \mathbf{q q'} | G (\Omega_+) | \mathbf{p p'} \rangle_A \, .
\label{eq:gmat}
\end{eqnarray}
where the single-particle spectra of nucleons in the BHF approach, $\ebhf$, are given by
\begin{eqnarray}
\ebhf = \ek + \Re \Sigma_{BHF}(k) \, .
\label{eq:sp_BHF}
\end{eqnarray}
In this approximation, the self-energy is given in terms of the $G$-matrix by:
\begin{eqnarray}
\Sigma_{BHF}(k) = \int \frac{\textrm{d}^3 k'}{(2\pi)^3} f[\varepsilon_{BHF}(k')]
\langle \mathbf{k k'} | G \left[ \Omega= \ebhf + \varepsilon_{BHF}(k') \right] | \mathbf{k k'} \rangle_A  \, .
\label{eq:se_BHF}
\end{eqnarray}
and the total energy per particle is obtained from:
\begin{eqnarray} 
\frac{E^{BHF}}{A} = \frac{\nu}{\rho} \int \frac{\mathrm{d}^3k}{(2\pi)^3}
                    \left[ \ek + \frac{1}{2} \Re \Sigma_{BHF}(k) \right] f[\ebhf]
\label{eq:ener_BHF}
\end{eqnarray}
We shall make two more comments concerning this finite temperature generalization
of the BHF approach. On the one hand, one must be aware that hole-hole propagation is
the cause of the Bose term in Eq.~(\ref{eq:sigma_ladder}). If the BHF is to be derived
from the SCGF approach, one should neglect this term, so that the imaginary part of
the self-energy reads:
\begin{eqnarray}
\textrm{Im }\Sigma_{BHF}(k) &=&
           \int \frac{\textrm{d}^3 k'}{(2\pi)^3} \, f[\varepsilon_{BHF}(k')]
           \langle \mathbf{kk}' | \Im G[\Omega = \ebhf+\varepsilon_{BHF}(k')]| \mathbf{kk}' \rangle_A \,
            .
\label{eq:sigma_BHF}
\end{eqnarray}
On the other hand, one should say that the BHF approach obtained from this
approximation of the SCGF method is not fully justified from basic first
principles. Indeed, the finite-temperature generalization of the $T=0$ Bethe-Goldstone
expansion is given by the BdD approach. In that theory, a similar expression to
Eq.~(\ref{eq:gmat}) for the in-medium temperature-dependent interaction
can be obtained, but in principle there is not a straightforward relation
with Eq.~(\ref{eq:ener_BHF}) for the energy per particle of the finite
temperature system. Nevertheless, one can see that this approximation is a reasonable
one, because 
the dominant diagrams of the BdD expansion reduce to this finite temperature
BHF approach at low temperatures \cite{baldo99}.
In the BHF approach that we will use here, the free energy is obtained
from the energy per particle of Eq.~(\ref{eq:ener_BHF})
together with the mean-field expression for the entropy of the system:
\begin{eqnarray}
S^{BHF} = - \intk \bigg\{ f[\ebhf] \ln f[\ebhf] + (1-f[\ebhf])
\ln(1-f[\ebhf])\bigg\} \, . \nonumber \\
\label{eq:s_BHF}
\end{eqnarray}

\subsection{Luttinger-Ward formalism}

\begin{figure}[thb]
   {\Large $\Sigma^{(n)}(k,\omega) = $
    \begin{minipage}{5cm} \vskip-0.4cm \includegraphics[width=5cm]{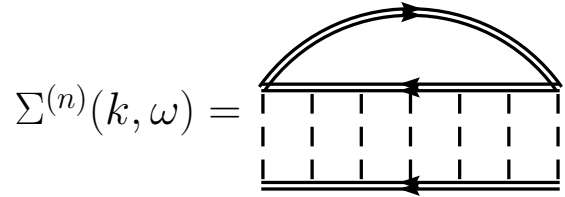} \end{minipage}}
    \vspace{0.75cm}
   \caption{$n$-th order contribution to the ladder self-energy.}
   \label{diag:sigma}
\end{figure}

\begin{figure}[thb]
   {\Large $\Phi[G] = \sum_n \frac{1}{2n} \bigg\{$ \hskip0.2cm  
           \begin{minipage}{4cm}   \includegraphics[width=4cm]{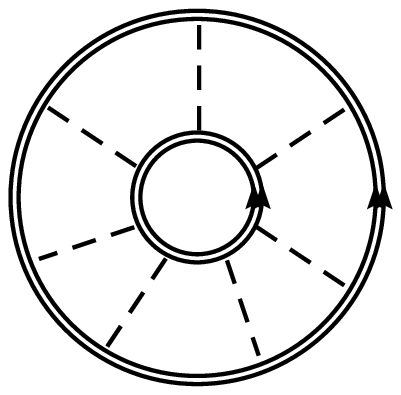} \end{minipage}
           $\bigg\}$ }
   \vspace{0.75cm}
   \caption{$\Phi$ functional for the $T$-matrix approximation.}
   \label{diag:phi}
\end{figure}

From a statistical mechanics point of view, the macroscopic information of the 
system is contained solely in the partition function. Once this function is known,
its derivatives give access to the TD properties of the system.
However, from a microscopic point of view, it is well known that the one-body
properties can be derived from the dressed single-particle propagator of the system.
Now one may ask whether there is a connection between both functions and, in
particular, if the single-particle propagator is enough for building the partition
function. The answer, given by Luttinger and Ward \cite{luttinger60} more than
forty years ago, is positive. The expression for the partition function thus
obtained turns out to have some interesting properties that were later on exploited
by Baym \cite{baym62} in his well-known discussion of the TD consistency of
many-body approaches. The starting point is the
so-called Luttinger--Ward expression of the partition function of the system:
\begin{eqnarray}
\ln Z &=& \Tr \Sigma(k,z_{\nu}) \, G(k,z_{\nu}) + \Tr \ln \big[ - G^{-1}(k,z_{\nu}) \big] 
		- \Phi \big[ G \big] \, ,
		\label{eq:log_Z}
\end{eqnarray}
where the functional $\Phi[G]$ has been introduced and where the trace $\Tr$ means a sum
over all momentum states and Matsubara frequencies:
\begin{eqnarray}
\Tr \rightarrow  \sum_{\vec{k},\nu} e^{z_{\nu} \eta} \, ,
\label{eq:trace}
\end{eqnarray}
with $\eta=0^+$ small and positive and such that $\lim_{\Re z \to \infty} \eta \Re z =\infty$.
In TD equilibrium, the grand-partition function is stationary under variations of the
Green's function:
\begin{eqnarray}
\left. \frac{\delta \ln Z}{\delta G} \right|_{G_0}=0 \, ,
		\label{eq:stat}
\end{eqnarray}
and thus the functional $\Phi$ should satisfy the following condition:
\begin{eqnarray}
\left. \frac{\delta \Phi}{\delta G} \right|_{G_0}= \Sigma \, .
		\label{eq:deltaphi}
\end{eqnarray}
To obtain the previous expression, we have used Dyson's equation, Eq.~(\ref{eq:dyson}) in
order to functionally derive the self-energy $\Sigma$ with respect to $G$. In fact, within the
LW formalism, $\Sigma$ is a functional of the
propagator and one can take Eq.~(\ref{eq:deltaphi}) as its definition.
Usually, however, in a given many-body approach one assumes a certain approximation
to the self-energy (in our case, the ladder approximation). The functional $\Phi$ is
then fixed by Eq.~(\ref{eq:deltaphi}) and the propagator is given by Dyson's equation,
$G^{-1} = G_0^{-1} + \Sigma$. These ideas can be depicted diagrammatically.
For the ladder approximation, we show in Fig.~\ref{diag:sigma} the $n$-th order
contribution to the self-energy $\Sigma^{(n)}$ (where $n$ denotes the number of
bare interaction lines in the diagram). By closing this diagram in its free vertices
with a propagator, we obtain the corresponding $n$-th order contribution to $\Phi$.
At each order, however, this contribution should be divided by a factor $2n$ that
takes into account the $2n$ possible places where one can cut each propagator
to obtain $\Sigma$ from Eq.~(\ref{eq:deltaphi}) \cite{baym62}.
The corresponding series for the functional is shown in terms of diagrams
in Fig.~\ref{diag:phi}.

The evaluation of the sum in the partition function has to be done with special
care because of the cut in the logarithm. The final expression for the
grand-potential $\Omega = - T \ln Z$ reads:
\begin{eqnarray}
\Omega &=& \intk \intw f(\omega) \, 2 \Im \bigg\{ \ln \big[ -G^{-1}(k,\omega_+) \big]
           + \Sigma(k,\omega_+) G(k,\omega_+) \bigg\} \nonumber \\
       &+& T \Phi [G] \, . 
             \label{eq:granpot}
\end{eqnarray}
In this expression, the propagator and the self-energy are computed above but
close to the real axis. We have also introduced the Fermi-Dirac distribution $f(\omega)$, 
Eq.~(\ref{eq:FD}). One can now readily obtain the entropy by means of the TD relation:
\begin{eqnarray}
S &=& - \left. \frac{\partial \Omega}{\partial T} \right|_{\mu} \, .
             \label{eq:s_TD}
\end{eqnarray}
The stationarity of $\Omega$ with respect to changes in $G$ is 
now very useful, because it implies that in Eq.~(\ref{eq:granpot}) 
only the temperature derivatives of the Fermi functions are needed:
\begin{eqnarray}
S &=& \intk \intw \frac{\partial f(\omega)}{\partial T} \, 2 \Im \bigg\{ \ln \big[ -G^{-1}(k,\omega_+) \big]
           + \Sigma(k,\omega_+) G(k,\omega_+) \bigg\} \nonumber   \\
  &+& \frac{\partial}{\partial T} T \Phi [G] \, .
             \label{eq:s_derT}
\end{eqnarray}
This expression gives the entropy per unit volume of a correlated
system of fermions as a function of $G$, $\Sigma$ and $\Phi$ and it
is the fundamental equation from which we will derive most of
our results. The usefulness and applications of this formula
in the context of Fermi-liquids were extensively discussed in the
pioneering work of Carneiro and Pethick \cite{carneiro75}. In the
following, we will closely follow this reference and discuss some of
the more relevant approximations for our case.

One can explicitly compute the imaginary parts of the two terms inside the 
integrals. The first term is the imaginary part of a logarithm, \emph{i.e.}
the phase of its argument. We can thus decompose the argument in its
real and imaginary parts:
\begin{eqnarray}
G^{-1}(k,\omega_+) &=& \omega  - \frac{k^2}{2m} - \Re \Sigma(k,\omega) - i \Im \Sigma(k,\omega_+) = 
                   \Re G^{-1}(k,\omega) + \frac{i}{2} \Gamma (k,\omega) \, .
\label{eq:gm1}
\end{eqnarray}
We consider that the logarithm has a cut in the real negative axis and we 
work in the sheet where $\ln (1) = 0$. The $\arctan(z)$ function goes from
$-\frac{\pi}{2}$ to $\frac{\pi}{2}$ and is normally used to obtain the phase of a
complex number. Nevertheless, whenever the real part of $G^{-1}$ becomes positive
and provided that $\Gamma(k,\omega)$ is positive defined, the complex
phase of the logarithm's argument will be in the III quadrant of the complex plane
and thus a factor $\pi$ needs to be subtracted to the the $\arctan$ function
in order to match the argument of $-G^{-1}$. The result
of the imaginary part can thus be casted in the following form:
\begin{eqnarray}
\Im \bigg\{ \ln \big[ -G^{-1}(k,\omega_+) \big] \bigg\} =
\arctan \lambda(k,\omega)
- \pi \theta \big[ \Re G^{-1}(k,\omega) \big] \, ,
             \label{eq:im_1}
\end{eqnarray}
where we have introduced the function $\lambda(k,\omega)$:
\begin{eqnarray}
\lambda(k,\omega) = \frac{ \Gamma(k,\omega)}{2 \Re G^{-1}(k,\omega)} \, ,
\label{eq:lambda}
\end{eqnarray}
which is nothing but the quotient between the imaginary and the real part
of the complex inverse propagator. 
On the other hand, the imaginary part of the second term in the integral of
Eq.~(\ref{eq:s_derT}) is given by:
\begin{eqnarray}
\Im \bigg\{ \Sigma(k,\omega_+) G(k,\omega_+) \bigg\} &=&
\Re \Sigma(k,\omega_+) \Im G(k,\omega_+) + \Im \Sigma(k,\omega_+) \Re G(k,\omega_+) \nonumber \\ 
&=& - \frac{1}{2} \Re \Sigma(k,\omega) \A(k,\omega) - \frac{1}{2} \Gamma(k,\omega) \Re G(k,\omega)\, .
             \label{eq:im_2}
\end{eqnarray}
Using Eqs.~(\ref{eq:im_1}) and (\ref{eq:im_2}), we can divide the entropy $S$ in two terms:
\begin{eqnarray}
S=S^{DQ} + S' \, ,         \label{eq:entropy}
\end{eqnarray}
given by:
\begin{eqnarray}
S^{DQ} &=& \intk \intw \frac{\partial f(\omega)}{\partial T} \, \Xi (k,\omega)
\label{eq:s_dq} 
\end{eqnarray}
and
\begin{eqnarray}
S' &=& - \frac{\partial}{\partial T} T \Phi [G]
+ \intk \intw \frac{\partial f(\omega)}{\partial T} \, \A(k,\omega) \Re \Sigma (k,\omega) \, ,
\label{eq:s_prime}
\end{eqnarray}
where we have introduced the function $\Xi(k,\omega)$:
\begin{eqnarray}
\Xi(k,\omega) &=& 2 \pi \theta \big[ \Re G^{-1}(k,\omega) \big]
- 2 \arctan \lambda(k,\omega) 
+ \Gamma(k,\omega) \Re G(k,\omega) \, .
\label{eq:xi}
\end{eqnarray}
$S^{DQ}$ is a dynamical quasi-particle (DQ) entropy which partially takes into account the
correlations of the dressed particles in the medium. It includes some finite width
effects, as seen by the fact that it is computed with a non-zero $\Gamma$. 
The term  $S'$, on the other hand, accounts for higher order correlations.
As it was shown in \cite{carneiro75}, this term arises from the cancellation between the
temperature derivative of $T \Phi$ and the second term of Eq.~(\ref{eq:s_prime}).
The only non-zero contributions that survive this cancellation come from terms in the
perturbation expansion that have at least two vanishing energy denominators.
In the following, we shall make the assumption that $S'$ is negligible. In this way,
our formalism is simplified because there is no need to evaluate the $\Phi$ functional.
This assumption, of course, needs to be validated, and that is what we will do in
the final part of this work.
As we will see, our approximation leads to TD consistent
results, which confirms that the contribution of $S'$ is
small in the density and temperature range explored
as far as short range correlations are concerned.
In addition, it is worth to mention that it is precisely from
the $S'$ contribution that all the anomalous temperature dependences
of the entropy arise. These anomalies, however, are mainly
generated by terms in $S'$ involving long-range correlations,
which we do not consider in our approach.
Thus, by restricting ourselves to computing the entropy with $S^{DQ}$,
we will loose these contributions and only find analytical ($S \sim T, T^3$) temperature
dependences.

To study the DQ entropy, it is interesting to analyze the
properties of the function $\Xi(k,\omega)$. The first term of $\Xi$ in Eq.~(\ref{eq:xi})
is a step function with
argument $\Re G^{-1}$. For a fixed momentum $k$ and as a function of the energy,
this argument is negative for energies below the quasi-particle peak $\eqp$,
whereas it is positive for $\omega$ greater than $\eqp$. Thus the step function
can be rewritten: 
\begin{eqnarray}
\Xi_1(k,\omega) &=& 2 \pi \theta \bigg[ \omega - \frac{k^2}{2m} - \Re \Sigma(k,\omega) \bigg] =
2 \pi \theta \big[ \omega - \eqp \big] \, .
\label{eq:xi1}
\end{eqnarray}
At a fixed momentum, then, $\Xi_1$ equals zero at energies below the
quasi-particle pole and $2 \pi$ above it. By using the relation:
\begin{eqnarray}
\frac{\partial f(\omega)}{\partial T} &=& - \frac{\partial \sigma(\omega)}{\partial \omega} \, ,
             \label{eq:difT}
\end{eqnarray}
where we have introduced the function:
\begin{eqnarray}
\sigma(\omega) = - \bigg\{ f(\omega) \ln f(\omega) + \big[ 1 - f(\omega) \big] \ln \big[ 1 - f(\omega) \big] \bigg\} \, ,
             \label{eq:sigma}
\end{eqnarray}
the contribution of $\Xi_1$ to the entropy is given by:
\begin{eqnarray}
S^{DQ}_1 &=& \intk \intw \frac{\partial \sigma(\omega)}{\partial \omega } \, 
2 \pi  \theta \big[ \omega - \eqp \big] =
\intk \int_{-\infty}^{\eqp} \mathrm{d} \omega \frac{\partial \sigma(\omega)}{\partial \omega } = \nonumber \\
&=& \intk \sigma \big[ \eqp \big] \equiv S^{QP} \, .
\label{eq:s_dq_1}
\end{eqnarray}
This expression corresponds to the entropy of a system of undamped quasi-particles 
with real quasi-particle energies given by Eq.~(\ref{eq:qp}). Whenever 
quasi-particles have long lifetimes, we expect it to be a good approximation to
the entropy. Indeed, for any many-body approximation where the quasi-particle
energies are real (such as the Hartree-Fock case, for instance) the full
DQ entropy is simply given by Eq.~(\ref{eq:s_dq_1}).
The rest of the terms in $\Xi$ can be rewritten as a function of $\lambda(k,\omega)$:
\begin{eqnarray}
\Xi_2(k,\omega) &=& - 2 \arctan \big[ \lambda(k,\omega) \big] \, ,
\end{eqnarray}
for the second term and:
\begin{eqnarray}
\Xi_3(k,\omega) &=& \frac{\lambda(k,\omega)}{1 + \lambda^2(k,\omega)} \, ,
\label{eq:xi_23}
\end{eqnarray}
for the third one. Their total contribution to the entropy is then given by:
\begin{eqnarray}
S^{DQ}_2 &=& \intk \intw \frac{\partial f(\omega)}{\partial T} \,
\bigg\{ \frac{\lambda(k,\omega)}{1 + \lambda^2(k,\omega)}
- 2 \arctan \big[ \lambda(k,\omega) \big] \bigg\} \,.
\label{eq:s_dq_2}
\end{eqnarray}
This expression involves a non-vanishing width $\Gamma$ and it can thus be
thought as a lifetime correction to the DQ entropy.
It is clear that for infinitely long lived quasi-particles ($\Gamma$=0),
this contribution will be zero, but for large widths it can have
a non-negligible effect on the total entropy. 

Let us go back to Eq.~(\ref{eq:s_dq}) for the DQ entropy. After a partial
integration and using relation (\ref{eq:difT}), we see that the
following expression for the DQ entropy holds:
\begin{eqnarray}
S^{DQ} &=& \intk \intw \sigma(\omega) \, \B(k,\omega) \, .
            \label{eq:s_dq_bsf}
\end{eqnarray}
provided that the $\B$ spectral function is defined as:
\begin{eqnarray}
\B(k,\omega) = \frac{\partial \Xi(k,\omega)}{\partial \omega} \, .
            \label{eq:bdef}
\end{eqnarray}
The expression Eq.~(\ref{eq:s_dq_bsf}) has several interesting properties. First of all,
in the free and the Hartree-Fock cases the $\B$ function reduces to a delta peak and $S^{DQ}$
becomes the expected expression for the entropy per particle. However, this does not
mean that $S^{DQ}$ neglects the finite width of quasi-particles,
as we already commented. In addition, it is easy to check that the $\B$ function
fulfils the following sum rule:
\begin{eqnarray}
\intw \B(k,\omega) = \frac{1}{2\pi} \Xi(k,\omega) \big|^{\infty}_{-\infty} = 1 \, .
           \label{eq:b_sumrule}
\end{eqnarray}
Finally, let us remark that this expression is somehow intuitive, in the sense that
it is a product of the statistical weighting factor for the entropy $\sigma(\omega)$
times a spectral function that takes into account the width of quasi-particles. 

An alternative way to obtain the entropy  within the LW formalism
is obtained by starting from Eq.~(\ref{eq:granpot}) and using the fact that:
\begin{eqnarray}
\beta f(\omega) = - \frac{\partial}{\partial \omega} \ln \left[ 1+ e^{-\beta(\omega-\mu)}\right] \, .
\label{eq:derf}
\end{eqnarray}
Integrating by parts one easily obtains: 
\begin{eqnarray}
\Omega &=& - \intk \intw T \ln \left[ 1+ e^{-\beta(\omega-\mu)}\right]
           B(k,\omega) + T \Phi [G] \, ,
             \label{eq:granpot2}
\end{eqnarray}
where we have introduced the $B$ function:
\begin{eqnarray}
B(k,\omega) &=& - \frac{\partial}{\partial \omega} 2 \Im \bigg\{ \ln \big[ -G^{-1}(k,\omega_+) \big]
                 + \Sigma(k,\omega_+) G(k,\omega_+) \bigg\} \, .
             \label{eq:bp1}
\end{eqnarray}
The entropy is now obtained via the temperature derivative of Eq.~(\ref{eq:granpot2}) which
will only affect the explicit temperature factors thanks to the stationarity
condition Eq.~(\ref{eq:stat}):
\begin{eqnarray}
S &=& \intk \intw \sigma(\omega) \, B(k,\omega) + \frac{\partial}{\partial T} T \Phi [G] \, .
             \label{eq:entro} 
\end{eqnarray}
Different expressions for $B$ can be obtained depending on whether the
derivative or the imaginary part are taken first. By taking first the imaginary
part and then the derivative, we see that the following relation holds between
$\Xi$ and $B$:
\begin{eqnarray}
B(k,\omega) &=& \frac{\partial}{\partial \omega} \Xi(k,\omega)
                + \frac{\partial \Re \Sigma(k,\omega)}{\partial \omega} \A(k,\omega)
                + \Re \Sigma(k,\omega) \frac{\partial \A(k,\omega)}{\partial \omega} \, .
            \label{eq:bp2}
\end{eqnarray}
For the first term of Eq.~(\ref{eq:bp1}), however, it is instructive to follow the
opposite direction, deriving first and taking after the imaginary part we obtain:
\begin{eqnarray}
B_1(k,\omega) &=& -2\Im \bigg\{ \frac{\partial}{\partial \omega}  \ln \big[ -G^{-1}(k,\omega_+) \big] \bigg\} =
-2\Im \bigg\{ G(k,\omega_+) \bigg[ 1 - \frac{\partial \Sigma(k,\omega_+)}{\partial \omega} \bigg] \bigg\} = \nonumber \\
&=& \A(k,\omega) \bigg[ 1 - \frac{\partial \Re \Sigma(k,\omega)}{\partial \omega} \bigg]
- \Re G(k,\omega) \frac{\partial \Gamma(k,\omega)}{\partial \omega} \, .
             \label{eq:bI}
\end{eqnarray}
The derivative of the second term is easily computed and can be separated in two parts:
\begin{eqnarray}
B_2(k,\omega) &=& \frac{\partial \Re G(k,\omega)}{\partial \omega} \Gamma(k,\omega)
                    + \Re G(k,\omega) \frac{\partial \Gamma(k,\omega)}{\partial \omega} \, ,
             \label{eq:bII}
\end{eqnarray}
and
\begin{eqnarray}
B_3(k,\omega) &=& \frac{\partial \Re \Sigma(k,\omega)}{\partial \omega} \A(k,\omega)
                    + \Re \Sigma(k,\omega) \frac{\partial \A(k,\omega)}{\partial \omega} \, .
             \label{eq:bIII}
\end{eqnarray}
Now we can write a compact expression for the $B$ function:
\begin{eqnarray}
B(k,\omega) &=& \A(k,\omega) +  \frac{\partial \A(k,\omega)}{\partial \omega} \Re \Sigma(k,\omega) 
                    + \frac{\partial \Re G(k,\omega)}{\partial \omega} \Gamma(k,\omega) \, .
             \label{eq:bfunction}
\end{eqnarray}
In addition, if one uses the fact that $B_3$ equals the last two terms of Eq.~(\ref{eq:bp2}),
it is easy to obtain the following relation between the $B$ and the $\B$ spectral functions:
\begin{eqnarray}
\B(k,\omega) &=& B_1(k,\omega) + B_2(k,\omega) =\A(k,\omega) \bigg[ 1 - \frac{\partial \Re \Sigma(k,\omega)}{\partial \omega} \bigg]
+ \frac{\partial \Re G(k,\omega)}{\partial \omega} \Gamma(k,\omega) \, ,
              \label{eq:b}
\end{eqnarray}
which also gives $\B$ as a function of $\A$ and the real and imaginary parts of the propagator.

The expression for the entropy of Eq.~(\ref{eq:entro}), in terms of the $\Phi$ functional plus
a term containing the integral of a statistical factor $\sigma(\omega)$ and the weighting
function $B(k,\omega)$ has already been obtained in, for instance, \cite{soma06}.
However, one should take
into account that our approximation for the entropy, $S^{DQ}$, differs from 
the first term of Eq.~(\ref{eq:entro}). In our case, we have neglected the terms of
Eq.~(\ref{eq:s_prime}) which are in fact canceling each other to a certain degree, whereas by
approximating the entropy with the term of the $B$ spectral function one would be ignoring
this cancellation. In fact, there is no reason to believe that the first term of
Eq.~(\ref{eq:entro}) should
be a good approximation to the full entropy,
while the DQ entropy, given by the convolution of the statistical factor $\sigma(\omega)$ and the
$\B$ spectral function, gives very reasonable results as we shall see in the following.

Also illustrative is the following decomposition of $S^{DQ}$ into two terms:
\begin{eqnarray}
S^A_1 &=& \intk \intw \sigma(\omega) \, \A(k,\omega) \label{eq:s_a1} \, ,
\end{eqnarray}
and
\begin{eqnarray}
S^A_2 &=& \intk \intw \sigma(\omega) \,
  \left\{ \frac{\partial \Re G(k,\omega)}{\partial \omega} \Gamma(k,\omega)
- \A(k,\omega) \frac{\partial \Re \Sigma(k,\omega)}{\partial \omega} \right\} \, .
            \label{eq:s_a2}
\end{eqnarray}
This justifies somehow the naive generalization to the expression of the entropy
that has been used in the literature \cite{cohen60,sedrakian06} and which consists in
approximating the entropy by formula (\ref{eq:s_a1}). This is, of course, not
justified from TD grounds, but it would be a natural extension of Eq.~(\ref{eq:nk})
to the case of the entropy. In particular, within a quasi-particle approximation,
when both $\A(k,\omega)$ and $\B(k,\omega)$ become quasi-particle delta functions,
$S^{DQ}$ and $S^A_1$ coincide and become the quasi-particle approximation to the
entropy, $S^{QP}$,
\begin{eqnarray}
S^{QP} &=&  \intk \sigma[\eqp] \label{eq:s_qp} \, .
\end{eqnarray}
The DQ entropy, however, goes beyond the naive quasi-particle
approach. It introduces the corrections of Eq.~(\ref{eq:s_a2}) which, as we
will see later on, are non-negligible.

Another justification for Eq.~(\ref{eq:s_dq_bsf}) comes from the generalization
of the $\Phi$-functional technique to non-equilibrium quantum systems. In
Ref.~\cite{ivanov00}, it was shown that, within certain $\Phi$-derivable 
approaches out of equilibrium, an $H$-theorem could be proved for a 
non-equilibrium kinetic entropy expressed in terms of the full Green's
function and the self-energy. When dealing with equilibrium systems, this
kinetic entropy reduces to the sum of $S^{DQ}$, the local or Markovian part
of the kinetic entropy ($S_{loc}$ in the language of Ref.~\cite{ivanov00}),
plus $S'$, the memory or non-Markovian part of the entropy ($S_{mem}$), which
coincides with the expression of the entropy Eq.~(\ref{eq:entropy}).

If $S^{DQ}$ is close to the real
entropy of the correlated system $S$, we should recover TD consistency,
\emph{i.e.}, the microscopically computed quantities should coincide
with the macroscopically computed ones. The ladder approximation is known to be a
well-defined $\Phi$-derivable approach \cite{baym62}. In particular, the
(microscopic) chemical potential $\tilde \mu$ computed from the normalization
condition:
\begin{eqnarray}
\rho = \intk \intw \A(k,\omega) f(\omega, \tilde \mu) \, ,
 \label{eq:norm} 
\end{eqnarray}
should coincide with the (macroscopic) chemical potential coming from the
TD expression:
\begin{eqnarray}
\mu (\rho,T) = \left. \frac{\partial F(\rho,T) }{\partial \rho} \right|_T \, ,
 \label{eq:mu} 
\end{eqnarray}
with $F$ the free-energy per unit volume. At $T=0$, this has been numerically
checked for the SCGF with separable potentials in~\cite{bozek01}. However,
it is also well known that some many-body approximations do not fulfil this
check of
consistency. The BHF approach, for instance, badly violates the Hugenholtz-van
Hove theorem \cite{hugenholtz58} which states that, at saturation density
the chemical potential $\tilde \mu$ and the free energy per particle $F/A$ should coincide.
The difference between these two quantities can be as large as $20$ MeV \cite{czerski02}.

In the following, we will present the numerical results for the $\B$ spectral
function and the different approximations to the entropy obtained from 
SCGF's calculations. Our aim is twofold. On the one hand, we will show that
the computation of the DQ entropy is enough to maintain
TD consistency; that is, that for the range of densities and temperatures
considered $S^{DQ}$ gives a free energy that respects the Hugenholtz-van Hove
theorem. This consistency is embedded in the ladder approximation (as shown by Baym \cite{baym62}),
but it is lacking in other many-body approaches.
In addition, the TD consistency of our results seems to indicate that the $S'$
contribution to the entropy is not crucial in the nuclear matter case.
Note
that an exact evaluation of $S'$ would need the knowledge of the $\Phi$-functional.
This functional has recently been computed also within a ladder approximation
\cite{soma06}, in a SCGF computation that differs from ours only in some numerical
details. The results that we will present (specially those concerning the
temperature dependence of the entropy) agree substantially with those of
Ref.~\cite{soma06}. We believe that this is an indication of the consistency of
both approaches.

\section{Microscopic results}

All the results quoted in this and the following sections have been obtained
with the finite temperature SCGF approach of Ref.~\cite{frick03} using the CDBONN
potential \cite{machleidt96}. In the numerical treatment, partial
waves up to $J=8$ have been included. The Born approximation has been used for $J \ge 3$.
The quoted BHF results have been computed with the same NN
potential with partial waves up to $J=4$. None of the calculations includes
three-body forces. We are thus not able to reproduce the saturation point of SNM.
In this sense, the results here presented should be taken as a first study
of the TD properties within SCGF theory, focused on the effects that
correlations induce on the entropy of SNM.

In the previous section we have mentioned that the properties of the $\B$ 
spectral function are very close to those of the usual spectral function
$\A(k,\omega)$. It fulfils a sum rule and it accounts somehow for the effect of
the width of quasi-particles in the DQ entropy. It is thus
natural to compare the two functions in the same plot. We can get a rough idea
of the differences of both functions following an argument first proposed by
Carneiro and Pethick. Let us express the spectral function $\A$ as a
function of the real and imaginary parts of the self-energy. We will of
course obtain the well-known Lorentzian-like function of
Eq.~(\ref{eq:asf_lor}).
For a given momentum, the spectral function will have a peak around the
quasi-particle energy of height $\A \sim 4/ \Gamma(k,\eqp)$. On the other
hand, the $\B$ spectral function can also be rewritten in terms
of the self-energy. Starting from Eq.~(\ref{eq:xi}) and taking
the derivative with respect to the energy, we get:
\begin{eqnarray}
\B(k,\omega) &=&
    \frac{1}{2} \frac{\Gamma^3(k,\omega)}{\left[ \left[ \omega - \ek - \Re \Sigma(k,\omega) \right]^2 +
                \left[ \frac{\Gamma(k,\omega)}{2} \right]^2 \right]^2}
                \left\{1 - \frac{\partial \Re \Sigma(k,\omega)}{\partial \omega} \right\} \nonumber \\
&-& \frac{1}{2} \frac{\Gamma^2(k,\omega)}{\left[ \left[ \omega - \ek - \Re \Sigma(k,\omega) \right]^2 +
                \left[ \frac{\Gamma(k,\omega)}{2} \right]^2 \right]^2}
                \bigg\{ \omega - \ek - \Re \Sigma(k,\omega) \bigg\}
                \frac{\partial \Gamma(k,\omega)}{\partial \omega} \, . 
             \label{eq:bsf_lor1}
 \end{eqnarray}
If one assumes that the frequency dependence of $\Gamma$ and $\Re \Sigma$ are
smooth close to the quasi-particle energy, we will have:
\begin{eqnarray}
\B(k,\omega) &\sim&
    \frac{1}{2} \frac{\Gamma^3(k,\omega)}{\left[ \left[ \omega - \ek - \Re \Sigma(k,\omega) \right]^2 +
                \left[ \frac{\Gamma(k,\omega)}{2} \right]^2 \right]^2} \, ,
             \label{eq:bsf_lor2}
 \end{eqnarray}
which corresponds to a function which decays faster than a Lorentzian close to $\eqp$,
but which has a stronger peak at the quasi-particle energy, $\B \sim 8/ \Gamma(k,\eqp)$.

\begin{figure}[thb]
   \includegraphics[height=18cm]{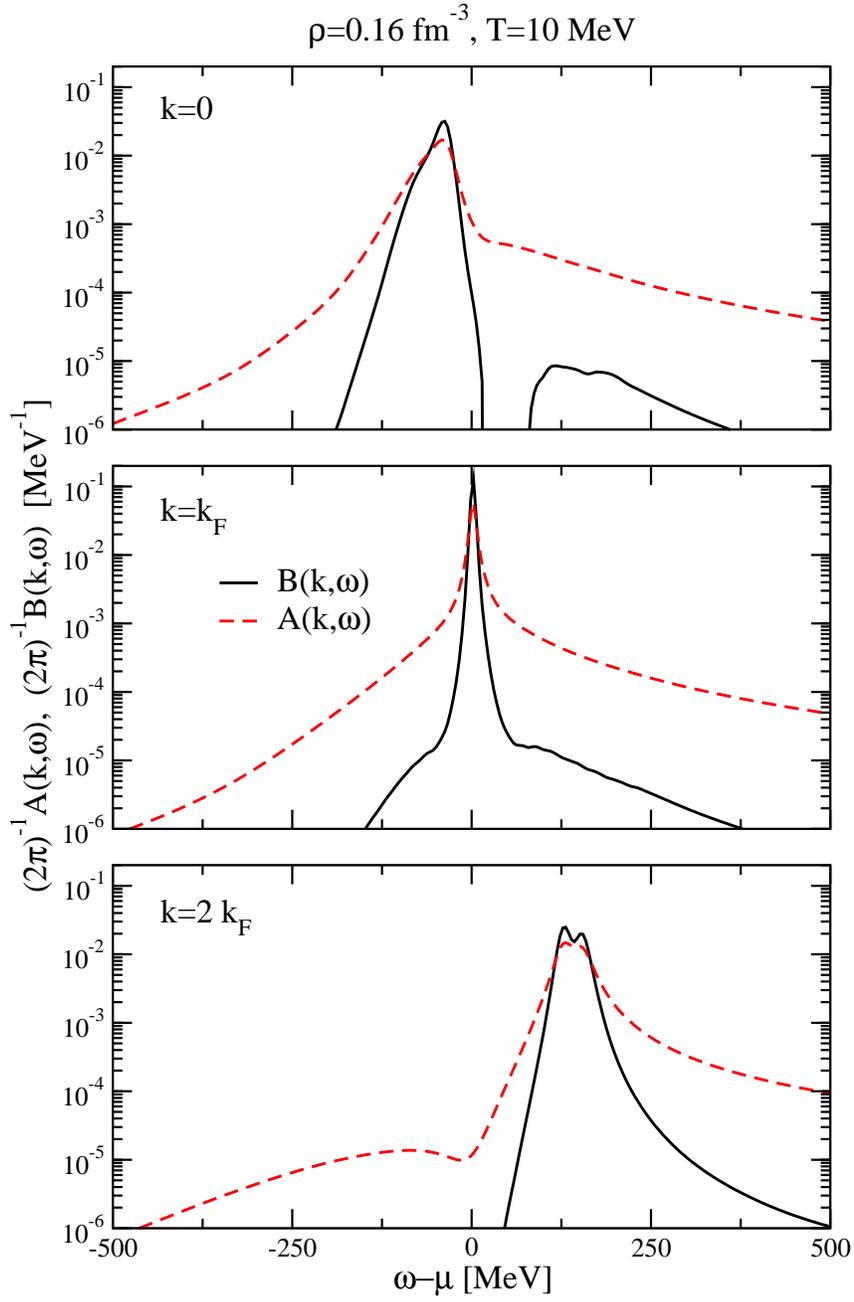}
   \vspace{0.75cm}
   \caption{(Color online) $\B$ (solid lines) and $\A$ (dashed lines) spectral functions at
   $\rho=0.16$ fm$^{-3}$ and $T=10$ MeV for three different momenta $k=0,k_F$ and $2k_F$.}
   \label{fig:a_vs_b}
\end{figure}

One can check that this schematic scenario is true in Fig.~\ref{fig:a_vs_b}, where
we show the $\B$ (full lines) and the $\A$ (dashed lines) spectral
functions as a function of the energy at three different momenta $k=0$, $k=k_F$ 
and $k=2k_F$, at the experimental saturation density $\rho=0.16$ fm$^{-3}$ and at a
temperature of $T=10$ MeV. In all the three panels, corresponding to the three
different momenta, we see that both functions are peaked around the same
energy values, corresponding to the quasi-particle energies given by Eq.~(\ref{eq:qp}).
The peaks shift from negative values (with respect to the chemical
potential) to positive values when going from zero-momentum to higher momentum 
states, just following the position of the quasi-particle peak. 
However, while the $\A$ spectral function has high-energy 
tails that contribute in a non-negligible way to the total strength of the nucleon,
the $\B$ spectral function has lower and less extended energy tails. This is 
easily understood if one considers that both functions fulfil the same
sum rule. Since the $\B$ function has a higher quasi-particle peak, the strength of
the peak is contributing substantially to the total sum rule and there is
no need to generate high-energy tails. The presence of these high-energy tails
in the $\A$ function is an indication of the importance of the correlations that 
go beyond the mean-field approach \cite{frick03}. Thus, the lack of such tails
in the $\B$ function is signaling somehow that these correlations will have a
small influence in the total entropy of the system.

This idea is also in accordance with the behaviour of the width of both spectral
functions. Far away from the Fermi momenta, both functions are relatively broad 
around the peak. Again, in the case of the $\A$ function this is a consequence of
the correlations that redistribute the nucleon single-particle strength within a wide
range of energies. The $\B$ function has a smaller width, which indicates that it is less
affected by correlations. Close to the Fermi momentum, however, both functions approach
a delta-peak behaviour, reminiscent of the fact that at zero temperature, even
when correlations are included, the spectral function has a delta-peak contribution.
At this momentum and for the temperature considered,
the $\B$ function is narrower and much more peaked than the
usual spectral function $\A$.

Also note that the values for the $\B$ function are positive for all the energies and
momenta that we have considered in our investigation. This is in contrast to
the weighting function $B$, which is defined in Eq.~(\ref{eq:bp1}) and has been
used in Ref.~\cite{soma06}. The fact, that the evaluation of the entropy using
the weighting function $B$ exhibits strong cancellation effects (see Fig.4 of 
Ref.\cite{soma06}) may be taken as an indication that the splitting of the
entropy into the two contributions according to Eq.~(\ref{eq:entro}) might not be
optimal.  

\begin{figure}[thb]
   \includegraphics[height=18cm]{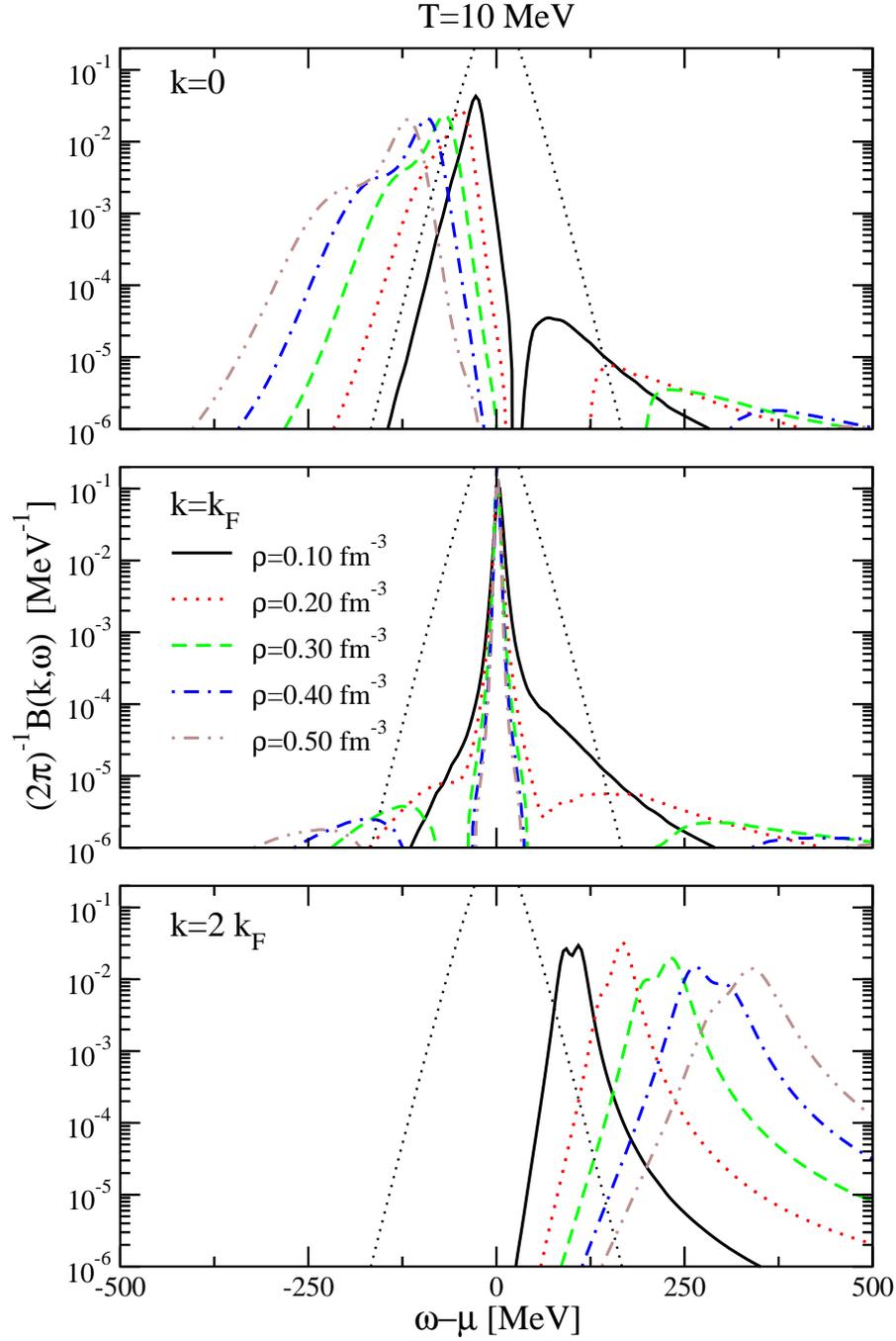}
   \vspace{0.75cm}
   \caption{(Color online) $\B$ spectral function for five different densities (from $\rho=0.1$ to $\rho=0.5$ fm$^{-3}$
    in equidistant steps) at a fixed temperature of $T=10$ MeV and three different momenta $k=0,k_F$ and $2k_F$.}
   \label{fig:brho}
\end{figure}

In order to understand the density dependence of the DQ entropy,
we show in Fig.~\ref{fig:brho} the $\B$ spectral function as a function of the energy
for different densities ($\rho=0.1, 0.2, 0.3, 0.4, 0.5$ fm$^{-3}$) at the same three
different momenta previously compared and at a fixed temperature of $T=10$ MeV.
In addition, we plot with a dotted line the statistical weighting function
$\sigma(\omega)$. It is precisely the product of these two functions, integrated over
energies and momenta, that gives rise to the DQ entropy, so
it is interesting to study their overlap.

The general features of the $\B$ spectral function as a function
of density are very
close to those of the usual spectral function $\A$. In the case of $k=0$, the quasi-particle
peak moves to more and more attractive energies as density increases, reflecting the
fact that the binding energy of a zero-momentum nucleon increases with density. 
Above the Fermi surface (at $k=2k_F$),
the situation is the opposite and the peak of the $\B$ function moves to higher
energies with increasing density.
The width of these peaks, both at zero momentum and at
twice the Fermi momentum, is broadened with density. This is
in accordance with the naive idea that correlation effects increase with density. In
addition, as a consequence of this broadening, the high energy tails (visible at
high positive energies for the $k=0$ state) decrease with density, allowing the sum
rule Eq.~(\ref{eq:b_sumrule}) to be fulfilled.

The situation is different at the Fermi surface: 
when the density is increased, the peak
remains at a fixed energy $\omega=\mu$, while its width becomes narrower and concentrates
more strength. This can
be understood if one takes into account that, as already commented,
at zero temperature, \emph{i.e.} for the fully degenerate system,
the correlated $\A$ spectral function shows a delta peak
which would also be present for the $\B$ spectral function.
At a fixed non-zero temperature, however, 
the system moves towards the degenerate limit (the ratio $T/\epsilon_F$
decreases) with increasing density and thus the $\B$ spectral function becomes closer to a delta peak.
This is actually what can be seen in the central panel of Fig.~\ref{fig:brho}. 
At high densities ($\rho \geq 0.2$ fm$^{-3}$), a clear separation between the quasi-particle peak and
the background contribution to the $\B$ spectral function is observed.

It is clear from Fig.~\ref{fig:brho} that the quasi-particle
peak and the peak of the $\sigma$ function only coincide for momenta close to $k_F$ and
energies around $\omega = \mu$. Thus,
the important contributions to the DQ entropy of the system will be that of
the momenta close to the Fermi surface and the energies close to the chemical potential.
It is precisely the interplay between $\sigma$ and $\B$ that
gives rise to the density dependence of the entropy. Since the value of $\B$ at
$k=k_F$ and $\omega = \mu$ increases with density, one may expect that the 
entropy per particle would increase with density. However, it is
also true that, for lower densities, the quasi-particle peak is closer to $\mu$ 
at all momenta and thus there are contributions of the quasi-particle peak for momenta
not necessarily close to $k_F$. In fact, when these contributions are summed up,
one finds that the entropy per particle decreases
with density.

\begin{figure}[thb]
   \includegraphics[width=14cm]{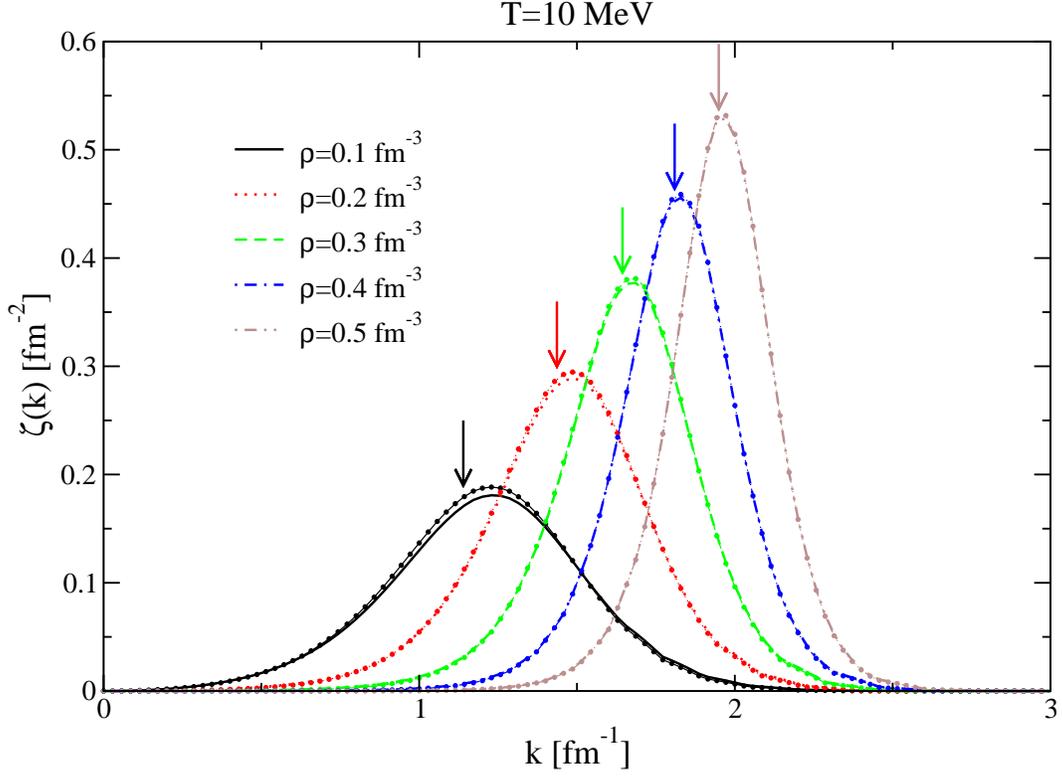}
   \vspace{0.75cm}
   \caption{(Color online) Momentum dependence of the $\zeta$ function (see
   Eq.~(\ref{eq:zeta})) for five different densities
   (from $\rho=0.1$ to $\rho=0.5$ fm$^{-3}$ in equidistant steps) at a fixed temperature of $T=10$ MeV.
   The dots correspond to the quasi-particle approximation $\zeta^{QP}$ for the same densities and temperature.
   The arrows signal the position of the Fermi momentum at each density.}
   \label{fig:zeta}
\end{figure}

To illustrate these results, we plot in Fig.~\ref{fig:zeta}, the momentum-dependent
integrand of Eq.~(\ref{eq:s_dq_bsf}):
\begin{eqnarray}
\zeta(k)=\frac{\nu}{2 \pi^2} k^2 \intw \sigma(\omega) \B(k,\omega) \, ,
\label{eq:zeta}
\end{eqnarray}
which measures the contribution to the DQ entropy from each momentum state.
It is clear that, as density increases, the integrand becomes larger at the
Fermi surface but less extended in momenta. This is in agreement with the previously
discussed ideas, \emph{i.e.}, that for less degenerate systems the contributions at
all momenta are relevant, while for degenerate systems the contribution of the
$k=k_F$ state is the most important one. In addition, the circles in the figure show
the $\zeta$ function obtained within a quasi-particle approximation:
\begin{eqnarray}
\zeta^{QP}(k)=\frac{\nu}{2 \pi^2} k^2 \sigma[\eqp] \, ,
\label{eq:zeta_qp}
\end{eqnarray}
with the quasi-particle peak given by the maximum of the spectral function,
Eq.~(\ref{eq:qp}). The differences are only relevant for the lowest densities and in
a range of momenta close to the Fermi momentum. This is again a signature of the small
role played by the correlations that deplete the quasi-particle states on the entropy.
Therefore, we expect that the quasi-particle approximation to the entropy, Eq.~(\ref{eq:s_qp}),
will describe the full DQ entropy $S^{DQ}$ quite well.

\begin{figure}[thb]
   \includegraphics[height=18cm]{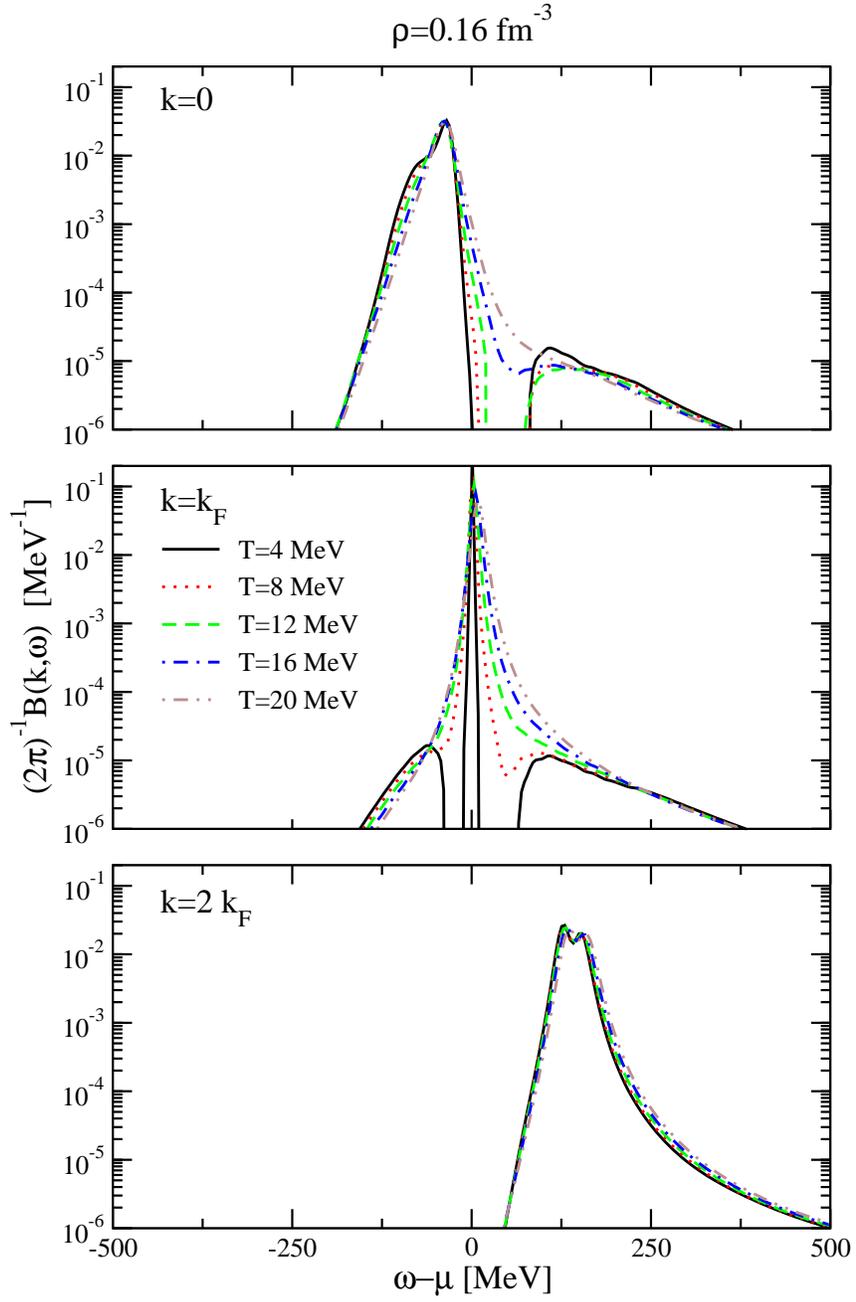}
   \vspace{0.75cm}
   \caption{(Color online) $\B$ spectral function for five different temperatures (from $T=4$ to $T=20$ MeV
   in equidistant steps) at a fixed density of $\rho=0.16$ fm$^{-3}$ and three different
   momenta $k=0,k_F$ and $2k_F$.}
   \label{fig:btemp}
\end{figure}

In order to gain insight in the temperature dependence of the DQ entropy, we show in
Fig.~\ref{fig:btemp} the $\B$ spectral function as a function of energy
for the same three different momenta considered previously at a fixed density $\rho=0.16$
fm$^{-3}$ and at 5 different temperatures $T=4,8,12,16,20$ MeV. It is clear that,
for all momenta, the variations of temperature mainly result in changes of the
width of the quasi-particle peak, while the position in energy of this peak
relative to the chemical potential is almost not changed. In addition, the momentum states
far above the Fermi
surface are not affected by temperature, as it is seen in the lowest
panel, corresponding to $k=2k_F$. At the Fermi surface, on the other hand, the
effects are much more important.
As temperature increases, the height of the quasi-particle peak decreases,
while its width increases.
Moreover, at the lowest temperature ($T=4$ MeV) a clean
separation is observed between a quasi-particle peak and a particle and hole background.
This separation is softened at $T=8$ MeV and disappears
completely above this temperature. Such a behaviour is again understood in terms
of the degeneracy of the system. The lower the temperature, the higher the
degeneracy and the smaller the width of the $\B$ spectral function.
For the $k=0$ state, a similar
situation is found. The peak lies below the chemical potential, and it is
clearly splitted from the particle background at $T=4$ MeV. For temperatures
above $T=12$ MeV this separation disappears and a smooth transition from particle
to hole states is found in the $\B$ function.
It is also interesting to notice that the width of the peak remains more
or less constant, thus indicating that temperature-induced effects on the
width of the quasi-particle peak are more important at $k=k_F$.

As for the total contribution to the DQ entropy, the convolution between
$\sigma(\omega)$ and $\B(k,\omega)$ is again crucial. At low temperatures,
$\sigma(\omega)$ is very peaked around $\omega \sim \mu$. The convolution
will thus only be different from zero whenever the quasi-particle peak is
close to $\mu$, \emph{i.e.}, at $k\sim k_F$. On the other hand, at higher
temperatures $\sigma(\omega)$ is different from zero in a wider region
of energies, which results in a non-zero convolution at all momenta.
When we integrate over momenta, the final DQ entropy is higher for the higher
temperature. Thus, in accordance with intuition, the entropy of this
correlated system will grow with temperature.

\section{Macroscopic results}

\begin{figure}[thb]
   \includegraphics[width=14cm]{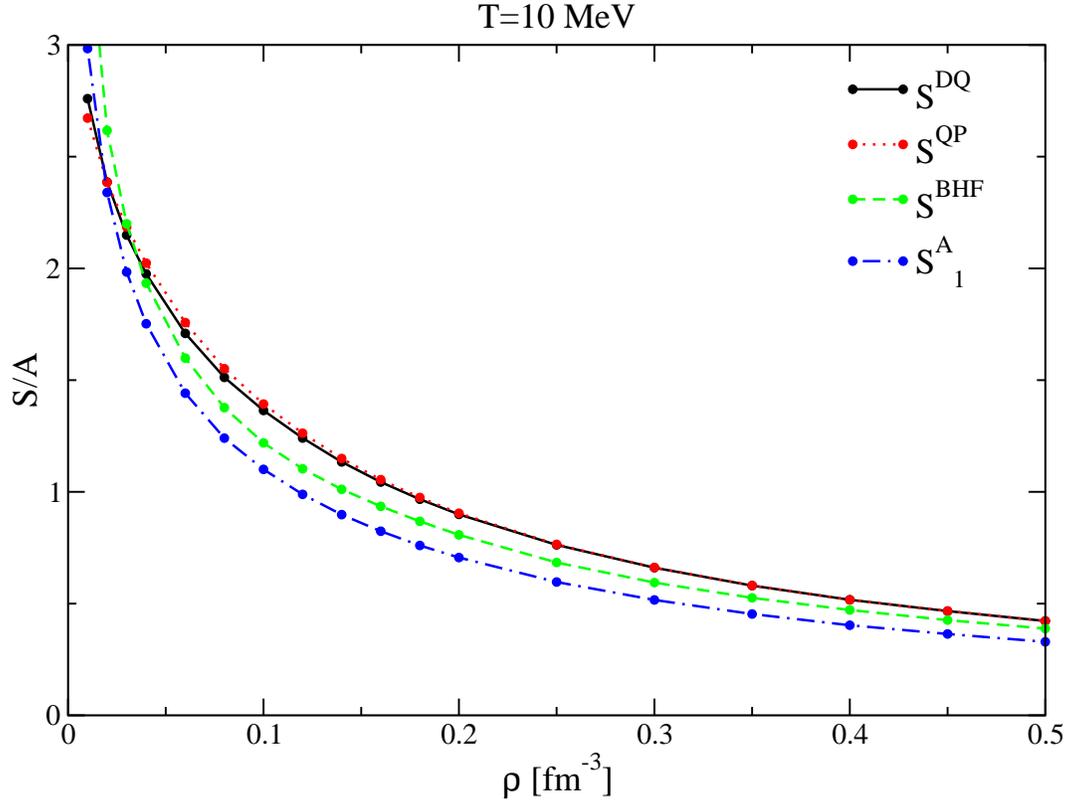}
   \vspace{0.75cm}
   \caption{(Color online) Different approximations to the entropy as a function of density for a $T=10$ MeV
   temperature. The full lines correspond to $S^{DQ}$; the dotted lines to $S^{QP}$;
   the dashed lines to $S^{BHF}$ and the dot-dashed lines to $S^{A}_1$.}
   \label{fig:srho}
\end{figure}

In this section we will explore the density and temperature dependences of
the entropy computed within different approximations. To begin with,
we show in Fig.~\ref{fig:srho} the density dependence of the entropy per
particle at a fixed temperature $T=10$ MeV. The approximations to the entropy
per particle that appear in this Figure are:
\begin{itemize}
\item $S^{DQ}$, the full dynamical quasi-particle entropy of Eq.~(\ref{eq:s_dq})
and equivalently Eq.~(\ref{eq:s_dq_bsf})
(solid lines).

\item $S^{QP}$, the quasi-particle approximation to the entropy of
Eq.~(\ref{eq:s_qp}) (dotted lines).

\item $S^{BHF}$, the Brueckner--Hartree--Fock entropy computed from expression
Eq.~(\ref{eq:s_BHF}) (dashed lines).

\item $S^A_1$, the contribution to the entropy due to the $\A$ spectral function
(dash-dotted line).
\end{itemize}
As a general feature, we can say that all of these entropies decrease substantially
with density, from values of around $2.5$ at densities of around $0.02$ fm$^{-3}$
down to values of around $0.4$ for the highest density here considered,
$\rho=0.5$ fm$^{-3}$.

One important result that arises from Fig.~\ref{fig:srho} is
the fact that, at all densities, $S^{DQ}$ and $S^{QP}$ are very close. This is
somehow in agreement with the idea that the inclusion of the width of
quasi-particles has a small effect in the entropy. As discussed in relation
with Fig.~\ref{fig:zeta}, the effect is larger at lower densities, where both
approximations differ more, but it is never higher than a $5 \%$.
At high densities, the difference is so small that it
cannot be appreciated in the Figure. This result is not at all
intuitive. It indicates that $S^{DQ}_2$, which is nothing but the difference
between $S^{DQ}$ and $S^{QP}$, decreases with density. But, since we have argued
that $S^{DQ}_2$ represents somehow the finite width effects on the
entropy and since correlations grow with density, we would also expect it to
grow with density. However, we have also seen that the higher the density, the smaller
the width of the $\B$ spectral function at $k=k_F$ (which
is the more relevant contribution at high densities) and thus the lower the
effects of correlations. This is why at higher densities both approximations to
the entropy tend to be similar. The fact that $S^{DQ}_2$ is negative at intermediate
densities (say from $\rho=0.05$ fm$^{-3}$ to $0.30$ fm$^{-3}$) is quite interesting:
in addition to stressing the fact that finite lifetime effects to the entropy
are small, one can say that it looks like correlations (\emph{i.e.} the width of the
quasi-particle) tend to order the system.

The effects of the self-consistent propagation of holes
are responsible for the difference between $S^{DQ}$ and
$S^{BHF}$. This can be clearly seen by rewriting Eq.~(\ref{eq:s_BHF}) to
give:
\begin{eqnarray}
S^{BHF} &=&  \intk \sigma[\ebhf] \label{eq:s_bhf2} \, ,
\end{eqnarray}
which differs from Eq.~(\ref{eq:s_qp}) in the position of the single-particle
peaks (given by $\eqp$ in the first case and by $\ebhf$ in the second)
and also in the different values of the chemical potentials in the
statistical factor $\sigma$. Since the effect of the width on $S^{DQ}$ is small
at this temperature, the difference between both entropies arises from the
different quasi-particle energies and chemical potentials of the two approaches.
Each of these quantities can differ by at most $20$ MeV
\cite{frick03}. However, the correction in the entropy is small probably due to
a cancellation between both differences in the argument of $\sigma$, where
$\varepsilon(k)$ and $\mu$ are subtracted.
In the intermediate
density region, the BHF entropy has values that are about $10$ \% below the
DQ one. The presence of hole-hole correlation, thus, increases the entropy,
\emph{i.e.} the thermal disorder. This is of course related to the fact that
hole-hole correlations tend to increase the density of single-particle states
close to the Fermi energy. If one tries to parameterize the quasi-particle
spectrum close to $\mu$ in terms of an effective mass $m^*$, one obtains 
larger values for the parameterization of the SCGF spectrum than for
BHF~\cite{frick03}.

Finally, we also show the contribution of $S^A_1$ to the DQ entropy. As we have
already mentioned, this expression comes from a naive generalization to
incorporate width effects, but
nevertheless it gives a reasonable first guess to the entropy per particle.
Intuitively, one would expect that, since the $\A$ spectral
function is wider than the $\B$ one, the overlap between $\A(k,\omega)$ and
$\sigma(\omega)$ at a given momentum should be higher and the final $S^A_1$ entropy would
overestimate $S^{DQ}$. However, this is not the case, except for the lowest
densities. This can be understood from the the height of the
quasi-particle peak for $\A$ being, roughly speaking, a factor of $2$ lower than
that for $\B$. Thus, although more extended in momentum, the $\zeta(k)$
function for the $\A$ spectral function is smaller and gives rise to a
lower entropy. The difference of both entropies is between  $20-30 \%$
for the intermediate density region.
The origin of such differences is the $S^A_2$
contribution of Eq.~(\ref{eq:s_a2}), which is the integral of two terms. Both
terms are of the same order at $\rho \sim 0.1$ fm$^{-3}$ but, while
the contribution proportional to $\A$ decreases with density, the one proportional
to $\Gamma$ increases, and above $\rho=0.3$ fm$^{-3}$ it carries more than
$80 \%$ of the total correction.

\begin{figure}[thb]
   \includegraphics[width=14cm]{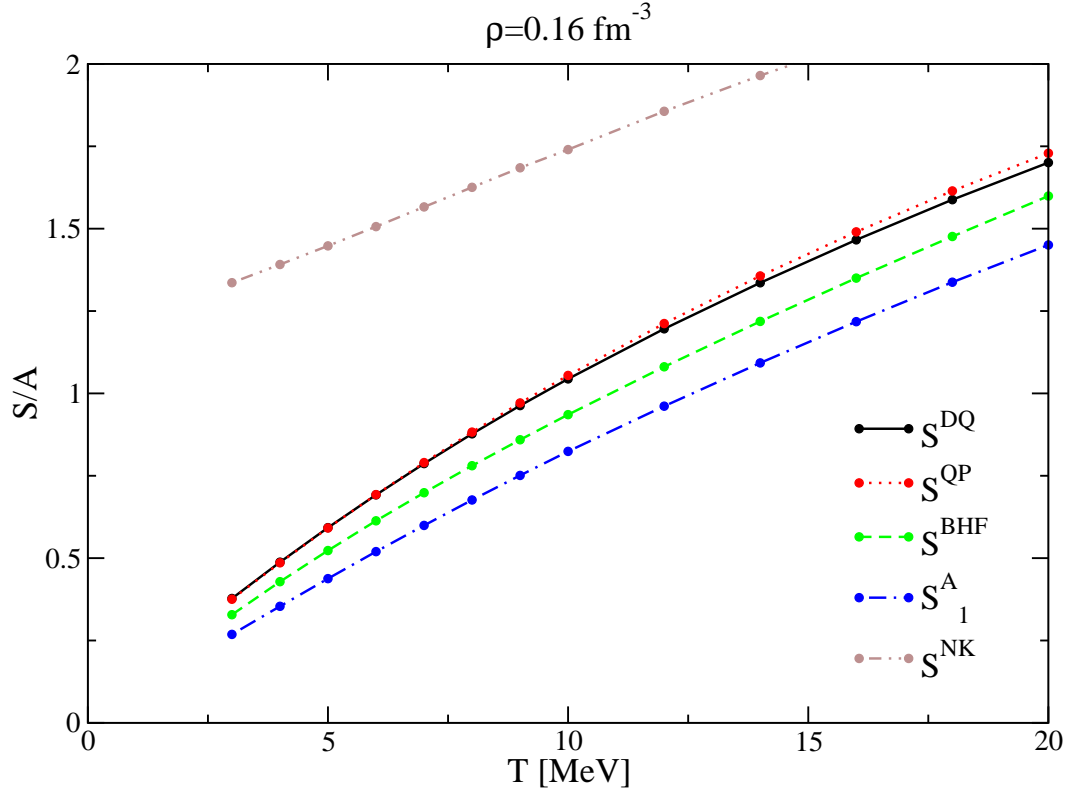}
   \vspace{0.75cm}
   \caption{(Color online) Different approximations to the entropy as a function of temperature at a density of
   $\rho=0.16$ fm$^{-3}$. The full lines correspond to $S^{DQ}$; the dotted lines to $S^{QP}$;
   the dashed lines to $S^{BHF}$; the dot-dashed lines to $S^{A}_1$ and the double dot-dashed lines
   to $S^{NK}$.}
   \label{fig:stemp}
\end{figure}

In Fig.~\ref{fig:stemp} we show the temperature dependence of the entropy for
a density $\rho=0.16$ fm$^{-3}$ computed with the same approximations as
discussed above. In addition, we have also computed another approximation, $S^{NK}$,
which is displayed with a double-dotted dashed line.
This corresponds to the entropy of the correlated momentum distribution, Eq.~(\ref{eq:nk}),
obtained from the mean-field-like expression:
\begin{eqnarray}
S^{NK} &=& - \intk \bigg\{ n(k) \ln n(k) + \big[ 1 - n(k) \big] \ln \big[ 1 - n(k) \big] \bigg\} \, .
 \label{eq:s_nk} 
\end{eqnarray}

One can say that all the approximations to the entropy of Fig.~\ref{fig:stemp}
decrease with temperature (as expected) and approach a linear dependence at low $T$.
It is a well known feature of Fermi liquids that the slope coefficient
for such a linear behaviour is proportional to the zero temperature density of
states computed at the Fermi surface, $N(T=0)$:
\begin{eqnarray}
\frac{S_{low}}{A} &=& \frac{\pi^2}{3 \rho} N(0) \, T \, .
 \label{eq:s_low} 
\end{eqnarray}
Each of our approximations goes to the $T=0$ limit with different slopes, and we
can thus obtain different densities of states. To calculate
$N(0)$ for each approximation, however, we should extrapolate our results to the $T=0$
limit, which we cannot do within our approach reliably. Instead,
from a low temperature expansion of expressions Eqs.~(\ref{eq:s_dq_bsf}) and
(\ref{eq:s_a1}) we obtain an expression of $N(0)$ in terms of the $\B$ and
$\A$ spectral functions that we extend to finite temperatures. Namely,
we get the ''density of states'' related to the $\B$ spectral function:
\begin{eqnarray}
N_{\B}(T) = \nu \int \frac{\mathrm{d}^3k}{(2\pi)^4} \, \B(k,\omega=\mu) \, ,
 \label{eq:N0_b} 
\end{eqnarray}
and the usual one, related to the $\A$ spectral function:
\begin{eqnarray}
N_{\A}(T) = \nu \int \frac{\mathrm{d}^3k}{(2\pi)^4} \, \A(k,\omega=\mu) \, ,
 \label{eq:N0_a} 
\end{eqnarray}
where $T$ denotes the fact that these have been computed at the finite temperature
at which the spectral functions have been computed.
Note that at low $T$ and $\omega = \mu$ the functions $\A$ and $\B$ differ basically
by a factor $\Z$ (where $\Z$ denotes the renormalization of the quasi-particle pole):
\begin{eqnarray}
\A(k,\omega=\mu) \sim \Z^{-1}(k,\omega=\mu) \B(k,\omega=\mu) \, 
 \label{eq:A_z_B} 
\end{eqnarray}
and thus the two densities of states will also differ by such a factor. Nevertheless,
the density of states that gives the correct linear fit to the DQ entropy is
that of Eq.~(\ref{eq:N0_b}). In fact, we have numerically checked that it is this
quantity that reduces to the well-known quasi-particle expression:
\begin{eqnarray}
N_{QP}(0) = \frac{\nu k_F m^*(k_F) }{2 \pi^2 \hbar^2} \,
 \label{eq:N0_qp} 
\end{eqnarray}
at low enough temperatures, where the effective mass is obtained through the derivative:
\begin{eqnarray}
\frac{m^*}{m} = \frac{1}{2m} \left( \frac{\textrm{d} \eqp}{\textrm{d} k^2}\right)^{-1} \,
 \label{eq:effm} 
\end{eqnarray}
evaluated at the Fermi surface using the finite temperature SCGF quasi-particle
spectrum, $\eqp$. This can be seen in Table~\ref{tab:n0}, where we give the densities
of states computed with the $\A$ and the $\B$ spectral functions 
together with that obtained from the QP expression, Eq.~(\ref{eq:N0_qp}), at
$\rho=0.16$ fm$^{-3}$ for low temperatures. The effective
mass at $k_F$ is given in the fifth column of Table~\ref{tab:n0}. The numerical values
confirm that at low temperatures the density of states from the $\B$ spectral function
reduces to the quasi-particle one, hence indicating that from a TD point of view this
is the correct density of states. However, from a
microscopic point of view, the $\A$ density of states is the one which has been
commonly used \cite{muther95,chen01}. One should keep in mind that
in a mean-field approximation both of them would reduce to the same expression,
Eq.~(\ref{eq:N0_qp}). 
It is also interesting to note that if we compute the entropy with the help of
the ``densities of states'' coming from the $\B$ spectral function, Eq.~(\ref{eq:N0_b}),
and from the QP expression, Eq.~(\ref{eq:N0_qp}) (both of them evaluated at the
corresponding finite temperature), and we use them instead of $N(0)$ in
Eq.~(\ref{eq:s_low}), we can reproduce the DQ expression of the entropy for
temperatures up to $T=10$ MeV with less than a $10$ \% discrepancy.

In the context of nucleus-nucleus collisions at intermediate energies, there exists
a growing amount of experimental data \cite{pochodzalla95,natowitz02}
which should be useful to
constrain the thermal properties of nuclei and nuclear matter. In
particular, the liquid-gas phase transition and the caloric curve give a hint on the
properties of nuclei at low temperatures. In the study of the caloric curve, it is
customary to parameterize the excitation energy at low temperatures in terms of the so-called
inverse level density parameter $K$ (see \cite{de98} for a theoretical description),
which is inversely proportional to the density of states introduced here.
The values we obtain for $K$ (defined as $K^{-1}=\frac{\pi^2}{6} N_{\B}(0)$) are close
to the Fermi gas value $K \sim 14.6$ MeV for $\rho=0.16$ fm$^{-3}$. This can
be understood from the fact that $N_{\B}$ reduces to the quasi-particle value of
Eq.~(\ref{eq:N0_qp}) which, in addition, is similar to the free Fermi gas value because
in our case the effective mass is almost equal to the bare nucleon mass at low temperatures.
A word of caution must be raised, however. Our value for $K$ is obtained from a calculation
in infinite isospin symmetric matter in which only short range correlations are treated.
Nevertheless, it is clear that a study of the inverse level parameter should
include both the effects of finite size and long-range correlations, which are very important in
determining the low energy excitations of nuclei.

In the following, we compare the different approximations to the entropy that appear in Fig.~\ref{fig:stemp}. As a general trend, we observe the same features that we have already discussed when we have commented the density dependence of the various approximations to the entropy.
The quasi-particle approximation
$S^{QP}$ using SCGF energies reproduces the DQ entropies at
all temperatures very well, especially below $T=10$ MeV. 
The finite temperature BHF
entropy describes the entropy of the system, with an error of
about $15$ \%. This difference is thus quite small, which is again a signal that
both the depletion of single-particle strength and the exact position of the
quasi-particle peak are not that crucial in the final result of the entropy per
particle. Let us also note that the inclusion of the hole-hole propagation
in the quasi-particle peaks tends to increase the entropy.

The $\A$ spectral function contribution to the entropy is again
correct in the temperature behaviour, but still it gives a too small value for the
entropy per particle of the system, with errors as large as a $30$ \%. As we have previously discussed, this is due to the lower quasiparticle peak of the $\A$ spectral function, which makes the $S^A_1$ contribution to the entropy lower than the $S^{DQ}$ entropy. The difference between these two functions is given by the $S^A_2$ contribution to the entropy. This is composed of the two terms in Eq.~(\ref{eq:s_a2}), which have
different relative weights as temperature changes. While the term proportional to
the $\A$ spectral function amounts to $90$ \% of the total correction at $T=4$ MeV,
its relative importance decreases linearly to a $30$ \% contribution for the
$T=20$ MeV case.

Finally, the entropy of Eq.~(\ref{eq:s_nk}) is given in terms of the fully correlated
momentum distributions. This momentum distribution includes both thermal effects
(which are capital for any entropy computation) plus correlation effects. In
fact, in the final entropy $S^{NK}$ both effects are taken as the same and
correlations somehow mimic extra thermal effects. This is why this is the only
approximation which tends to give a non-zero entropy at $T=0$. In the fully degenerate
limit, the momentum distributions given by Eq.~(\ref{eq:nk}) are not
Fermi step functions and they are corrected by correlation effects. These correlations
are the responsible for a certain amount of entropy, when this is computed with
Eq.~(\ref{eq:s_nk}) at $T=0$.
Thus, at finite temperatures thermal effects are overestimated in $S^{NK}$ due to the
presence of correlations, and $S^{NK}$ produces a far too large entropy (almost a
factor of three too large at $T=5$ MeV).

\begin{figure}[thb]
   \includegraphics[width=14cm]{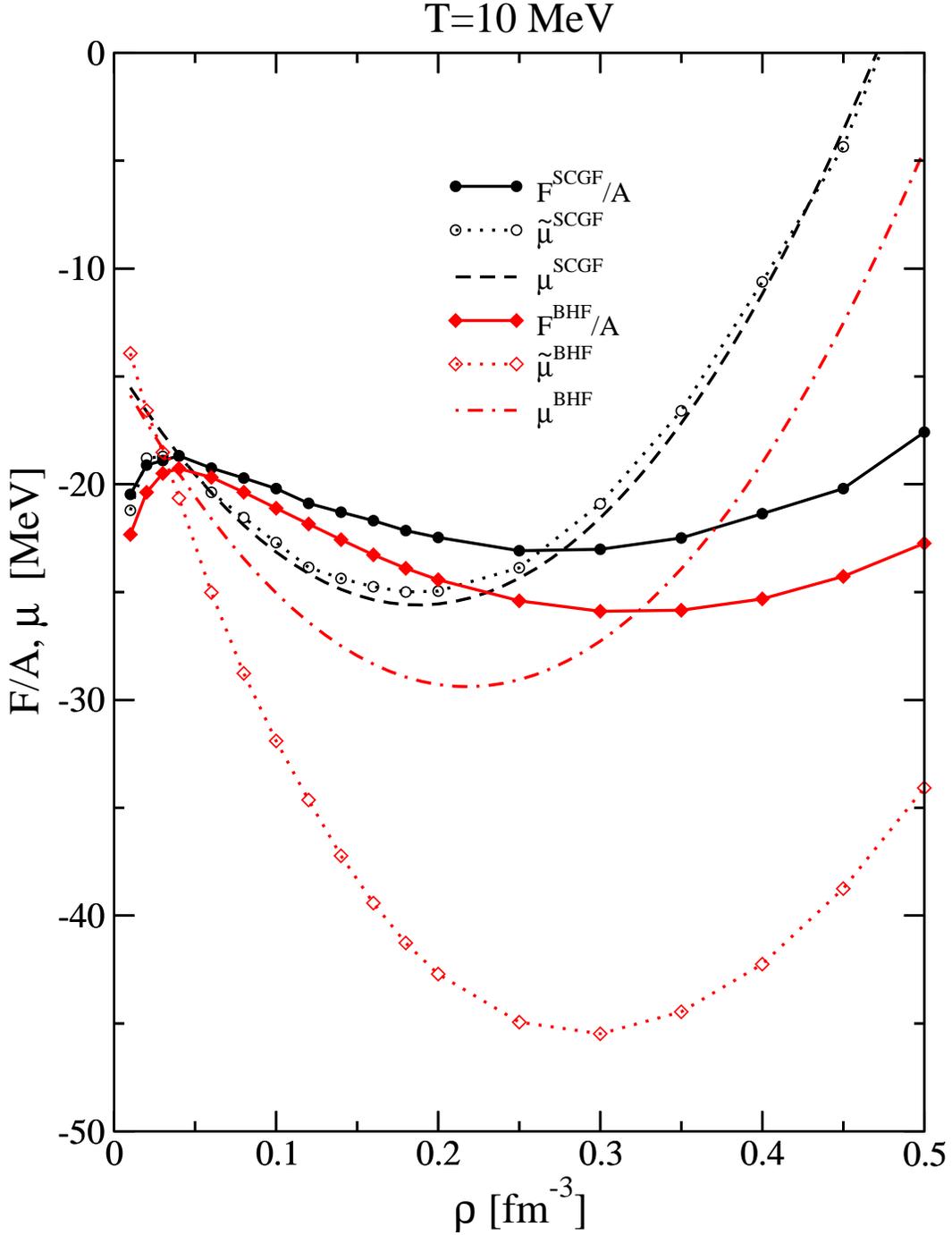}
   \vspace{0.75cm}
   \caption{(Color online) Free energies per particle (full lines) and $\tilde \mu$ chemical potentials (dotted lines)
   of the SCGF (circles) and BHF (diamonds) approaches as a function of density at a temperature of
   $T=10$ MeV. The $\mu$ chemical potential obtained through a numerical derivative are displayed in
   a dashed line for the SCGF results and a dot-dashed line for the BHF results.}
   \label{fig:TD}
\end{figure}

After having computed the entropy, we would like to address the subject of TD
consistency. From first principles, the ladder approximation is known to be
$\Phi$-derivable \cite{baym62}. Indeed, we have computed an expression
for the entropy within a formalism that preserves $\Phi$-derivability. Thus,
whenever $S'$ is negligible in our approach, we expect our results to preserve
TD consistency. Figure~\ref{fig:TD} shows the accuracy that we reach with
our SCGF results. For the sake of comparison, we also show the BHF results.
The upper solid lines with full circles correspond to the free energies per
particle computed within the SCGF approach:
\begin{eqnarray}
F^{SCGF} = E^{GMK} - T S^{DQ} \, ,
 \label{eq:F_SCGF} 
\end{eqnarray}
with the energy computed with the GMK sum rule Eq.~(\ref{eq:koltun}) and
the entropy with the DQ expression Eq.~(\ref{eq:s_dq}).
The BHF free energy is shown in a full line with solid diamonds and is
simply given by:
\begin{eqnarray}
F^{BHF} = E^{BHF} - T S^{BHF} \, ,
 \label{eq:F_BHF} 
\end{eqnarray}
with the energy computed from the generalization of the $T=0$ BHF approach
Eq.~(\ref{eq:ener_BHF}) and the entropy from Eq.~(\ref{eq:s_BHF}). The dotted
lines with empty circles and diamonds correspond to the microscopic chemical
 potentials $\tilde \mu$ obtained from inverting Eq.~(\ref{eq:norm}) for
 the SCGF and from inverting:
\begin{eqnarray}
\rho = \intk f(\ebhf, \tilde \mu^{BHF})
 \label{eq:norm_BHF}
\end{eqnarray}
in the case of BHF, respectively. Both chemical potentials are compared
with the macroscopic chemical potentials, $\mu$, obtained from the derivatives
of the free energies with respect to density Eq.~(\ref{eq:mu}), shown with a
dashed line for the SCGF and by a dashed-dotted line for SCGF. The derivative
has been performed numerically after adjusting $F$ to a third-order polynomial.
Although the low density region is not
well reproduced in this rough approximation, the results of the intermediate
density region can be fully trusted and in addition they are smooth with density.

The fulfilment of TD consistency for the SCGF approach is nicely illustrated in
Fig.~\ref{fig:TD}. The agreement between $\tilde \mu$ and
$\mu$ is very good above $0.05$ fm$^{-3}$, with discrepancies of less than 1
MeV up to $ \rho=0.5$ fm$^{-3}$. As a consequence, the Hugenholtz-van Hove
theorem is also very well fulfilled, and the minimum of $F/A$ and $\tilde \mu$
do nicely coincide at about $\rho \sim 0.27$ fm$^{-3}$. The situation for the
BHF approach, on the other hand, is much worse, as it is very well known
\cite{czerski02}. The chemical potentials
$\tilde \mu$ and $\mu$ differ by about $10$
MeV at $\rho=0.16$ fm$^{-3}$ and by almost $30$ MeV at the
highest density here considered. In addition, the Hugenholtz-van Hove theorem
is badly violated, and the value of $F/A$ at saturation differs from $\tilde \mu$
 by about $20$ MeV. Finally, we note that the propagation of holes
seems to have a global repulsive effect on the free-energy. Such a repulsive
effect has
already been detected for the energy per particle \cite{frick03}.
It is interesting to note, however, that the differences in free energies between
the BHF and the SCGF are smaller than the differences in energies, due to the
different entropy contributions.

The results presented here should be taken as a first step towards a full treatment
of the thermal properties of infinite nuclear matter within a formalism that includes
short-range correlations in a TD consistent way. From our point of view, this formalism
can find several applications within the many-body and nuclear physics community.
A first outcome could be, for instance, the study of the liquid-gas phase transition of
symmetric nuclear matter from a realistic NN potential. A critical study of the
usually assumed low-temperature dependences for the relevant TD properties of matter (of the
type $T^2$) could also be assessed. Other applications, such as the
study of the finite temperature equation of state, will probably demand for an inclusion
of three-body forces in the formalism in order to reproduce the empirical
saturation properties of nuclear matter. As for the possible consequences of our results for
the study of intermediate energy heavy ion collisions \cite{danielewicz02}, we expect that
they could affect the analysis based on transport models. The use of spectral functions in the description of correlated nucleons goes beyond
the quasi-particle picture customarily used in transport codes.
Although some results point out the fact that off-shell
effects in the propagation of nucleons are small \cite{cassing00}, a treatment of
a kinetic equation including full spectral functions obtained from realistic NN potentials
(following, for instance, \cite{botermans90} or \cite{kohler95}) is, to our knowledge, still
lacking. Moreover, even within the usual quasi-particle description, some of the in-medium
modifications of nucleons (such as effective masses or NN cross-sections) are taken as
simple parameterizations \cite{danielewicz00}. Our model permits the calculations of
these quantities from realistic NN potentials in a fully microscopical and TD consistent
basis which, properly parameterized, could be used in this kind of studies. 
                                                                         
\section{Summary and conclusions}

The ladder or $T$-matrix approximation of the SCGF method is a $\Phi$  derivable
approach and thus theoretically it should fulfil TD consistency. We have
checked the numerical TD consistency of this approach for symmetric nuclear
matter  at finite temperature by computing an approximation to the entropy, the
dynamical quasi-particle entropy, $S^{DQ}$, which has been discussed by Carneiro
and Pethick \cite{carneiro75}. Using this $S^{DQ}$ approximation we obtain a good
agreement between the chemical potentials determined within the SCGF calculation
and the corresponding values derived from the thermodynamical relations.
Therefore, the ladder approximation of the SCGF approach supplemented by the
evaluation of the entropy $S^{DQ}$ provides a method of calculating
thermodynamic properties of nuclear matter, which accounts for correlations
beyond the mean field picture in a consistent way.  

The entropy, $S^{DQ}$, can be evaluated in terms of a weighting function $\B$,
which is connected to the usual single-particle spectral function $\A$ of the
SCGF. This means that $\B$ can be calculated in term of the nucleon self-energy
and an explicit evaluation of the generating functional $\Phi$ can be avoided.
The $\B$ spectral function and its
momentum, energy, density and temperature dependences have been studied. 
In general, one can say that this function exhibits a more pronounced
quasi-particle structure than the corresponding spectral function $\A$.
As a consequence, the correlation effects related to the broadening of the
quasi-particle peak are not very important and a quasi-particle approximation
to the evaluation of the entropy is a very good approximation if the
quasi-particle energies are derived from the SCGF self-energy. 
Even if the quasi-particle energies are approximated by the single-particle
energies derived from the temperature dependent BHF approximation, the values
for the calculated entropy deviate only by as much as 10 to 20 percent.
In contrast to the SCGF, however, the BHF approximation fails to fulfil
thermodynamic consistency. The microscopic and the thermodynamic chemical
potential deviate substantially in the case of BHF and the Hugenholtz-van Hove
theorem is violated by more than 20 MeV. 

\section*{Acknowledgments}

The authors are very grateful to Dr. T. Frick, Dr. J. Margueron and Dr. A.
Sedrakian for useful and stimulating discussions. 
Arnau Rios acknowledges the support of DURSI and the European Social Funds.
This work was supported by the Grants No. FIS2005-03142 (MEC, Spain and Feder)
and No. 2005SGR-00343 (Generalitat de Catalunya).


\newpage
\begin{table}
\begin{displaymath}
\begin{array}{c|c|c|c|c|}
T \textrm{ [MeV]} & N_{\A}(T) \textrm{ [MeV$^{-1}$fm$^{-3}$]} & N_{\B}(T) \textrm{ [MeV$^{-1}$fm$^{-3}$]} & N_{QP}(T) \textrm{ [MeV$^{-1}$fm$^{-3}$]} & \frac{m^*}{m} \\ \hline
4 & 0.00435 & 0.00608 & 0.00608 & 0.935 \\
6 & 0.00430 & 0.00585 & 0.00586 & 0.901 \\
8 & 0.00424 & 0.00566 & 0.00570 & 0.875 \\
10 & 0.00416 & 0.00548 & 0.00557 & 0.855 \\
\end{array}
\end{displaymath}
\caption{Densities of states related to the $\A$ (first column) and $\B$ (second column) spectral functions 
at $\rho=0.16$ fm$^{-3}$ for different temperatures. The quasi-particle approximation to the density of
states, Eq.~({\ref{eq:N0_qp}}), is displayed in the fourth column, together with the effective mass at the
Fermi surface in the fifth column.}
      \label{tab:n0}
\end{table}

\newpage

\newpage

\newpage

\newpage

\newpage

\newpage

\newpage

\newpage

\end{document}